\documentclass[12pt,a4paper]{book}
\usepackage[utf8]{inputenc}
\usepackage{amsthm,amssymb,euscript,latexsym}
\usepackage[all,2cell]{xy}
\usepackage[ruled,vlined,linesnumbered]{algorithm2e}
\usepackage{hyperref}
\hypersetup{linkbordercolor=1 1 1}
\usepackage{anysize} 
\marginsize{3cm}{2cm}{2.5cm}{2.5cm}
\usepackage{fancybox, calc} 
\usepackage{oz}
\usepackage[dvips,pdftex]{graphicx}

\newcounter{nivell}
\makeatletter
\def\cleardoublepage{\clearpage\if@twoside \ifodd\c@page\else
  \hbox{}
  \vspace*{\fill}
  \vspace{\fill}
  \thispagestyle{empty}
  \newpage
  \if@twocolumn\hbox{}\newpage\fi\fi\fi}
\makeatother

\makeatletter
\def\thickhrulefill{\leavevmode \leaders \hrule height 1pt\hfill \kern \z@}
\def\s@btitle{\relax} 
\def\subtitle#1{\gdef\s@btitle{#1}} 
\renewcommand{\maketitle}{\begin{titlepage}%
    \let\footnotesize\small
    \let\footnoterule\relax
    \parindent \z@
    \reset@font
    \null\vfil
    \hrule height 1ex
    \vskip 10\p@
    {\raggedright
      \LARGE 
      \strut \@title \par}
      \vskip 20\p@
     {\raggedleft\LARGE \strut \@author\par}
    
    \vskip 5\p@
    \hrule height 1ex
    \vskip 40\p@
    \vfil\null
  \end{titlepage}%
  \setcounter{footnote}{0}%
}
\makeatother
\makeatletter
\def\thickhrulefill{\leavevmode \leaders \hrule height 1ex \hfill \kern \z@}
\def\@makechapterhead#1{%
  \vspace*{10\p@}%
  {\parindent \z@ \centering \reset@font
        \thickhrulefill\quad
        \scshape \@chapapp{} \thechapter
        \quad \thickhrulefill
        \par\nobreak
        \vspace*{10\p@}%
        \interlinepenalty\@M
        \hrule
        \vspace*{10\p@}%
        \Huge \bfseries #1\par\nobreak
        \par
        \vspace*{10\p@}%
        \hrule
    \vskip 40\p@
  }}
\def\@makeschapterhead#1{%
  \vspace*{10\p@}%
  {\parindent \z@ \centering \reset@font
        \thickhrulefill
        \par\nobreak
        \vspace*{10\p@}%
        \interlinepenalty\@M
        \hrule
        \vspace*{10\p@}%
        \Huge \bfseries #1\par\nobreak
        \par
        \vspace*{10\p@}%
        \hrule
    \vskip 40\p@
  }}

\theoremstyle{plain}
\newtheorem{theorem}{Theorem}
\newtheorem{proposition}{Proposition}
\newtheorem{problem}{Problem}
\newtheorem{lema}{Lemma}

\theoremstyle{definition}

\newtheorem{definition}{Definition}
\newtheorem{example}{Example}

\theoremstyle{remark}
\newtheorem{obs}{Remark}
\newtheorem{observation}{Remark}
\newtheorem{note}{Note}
\newtheorem{remark}{Remark}
\newtheorem*{Proof}{Proof}

\usepackage{alltt, fancyhdr, boxedminipage}
\usepackage{makeidx, multirow, longtable, tocbibind, amssymb}

\newlength{\funcindent}
\newlength{\funcwidth}
\newlength{\varindent}
\newlength{\varnamewidth}
\newlength{\vardescrwidth}
\newlength{\varwidth}
 {\begin{list}{}{%
   \settowidth{\labelwidth}{\texttt{#1:}}%
   \setlength{\leftmargin}{\labelsep}%
   \addtolength{\leftmargin}{\labelwidth}}}%
 {\end{list}}
\usepackage{enumerate}
\usepackage{array}
\usepackage{makeidx}

\author{Adrià Alcalá Mena}
\title{Trivalent Graph isomorphism in polynomial time}
\date{\today}
\makeindex

\begin{document}
\thispagestyle{empty}
\vspace*{7mm}
\begin{center}
{\Large\bf TRIVALENT GRAPH ISOMORPHISM }

{\Large\bf IN POLYNOMIAL TIME }\\[5mm]
\hrule

\vspace{10mm}
{\Large Facultad de Ciencias}\vspace*{3mm}

{\Large Universidad de Cantabria}

\vspace{10mm}

\includegraphics[width=25mm,height=25mm]{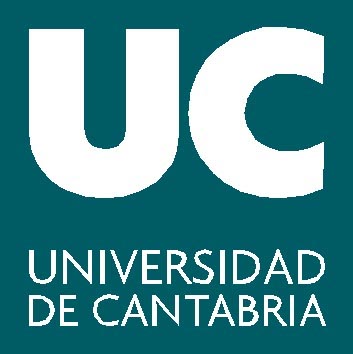}

\vspace{10mm}

{\large Programa Oficial de Postgrado de
Ciencias, Tecnolog\'ia y Computaci\'on}

{\large Máster en Matem\'aticas y Computaci\'on} \vspace{3mm}

{\large Master Thesis}


{\large\bf Adrià Alcalá Mena}\\[1cm]



{\large Junio 2012}\\[2cm]

\end{center}\vfill

\hfill{\large Advisor: {\bf Jaime Guti\'errez}}

 \newpage

\tableofcontents
 \newpage

\chapter*{Preamble}
\addcontentsline{toc}{chapter}{Preamble}

The graph isomorphism problem has a long history in mathematics and computer science, and more recently in 
fields of chemistry and biology. Graph theory is a branch of mathematics started by Euler as early as 1736 with his paper 
\textit{The seven bridges of Könisberg}. It took a hundred years before other important contribution of Kirchhoff had been 
made for the analysis of electrical networks. Cayley and Sylvester discovered several properties of special types of graphs 
known as trees. Poincaré defined  what is known nowadays as the incidence matrix of a graph. It took another 
century before the first book was published by Dénes Kőnig at 1936 titled \textit{Theorie der endlichen und unendlichen Graphen}. 
After the second world war, further books appeared on graph theory, for example the books of Ore, Behzad and Chartrand, 
Tutte, Berge, Harary, Gould, and West  among many others.

The graph isomorphism problem is the computational problem of determining whether two finite graphs are isomorphic. Besides it's 
practical importance, the graph isomorphism problem it's one of few problems which belonging to NP neither known to be solvable 
in polynomial time nor NP-complete. It is one of only 12 such problems listed by Garey $\&$ Johnson(1979), and one of only two 
of that list whose complexity remains unresolved (the other being integer factorization). It is known this computational problem
is in the low hierarchy of class NP, which implies that it is not NP-complete unless 
the polynomial time hierarchy collapses to its second level. Since the graph isomorphism problem is neither known to be NP-complete nor to be tractable, researchers have  sought to gain insight into the problem by defining a new class GI, the set of problems with a polynomial-time Turing 
  reduction to the graph isomorphism problem \cite{Booth77problemspolynomially}. In fact, if the graph isomorphism problem is solvable in polynomial time, 
then  GI would equal P.

  The best current theoretical algorithm is due to Eugene Luks (1983) and is based on the earlier work by Luks (1981), Babai and Luks (1982), combined with a subfactorial algorithm due to Zemlyachenko (1982). The algorithm relies on the  classification of finite simple groups, 
  without these results a slightly weaker bound $2^{O(\sqrt{n}\log^2 n)}$ was obtained 
  first for strongly regular graphs by László Babai (1980), and then extended to general graphs by Babai and Luks (1982), where $n$ is the number of the vertices.
  Improvement of the exponent $\sqrt{n}$ is a major open problem; for strongly regular graphs this was done by Spielman 
  (1996).

There are several practical applications of the graph isomorphism problem, for example, in chem-informatics and in mathematical chemistry; 
graph isomorphism testing is used to identify a chemical compound within a chemical database. Also, in organic mathematical 
chemistry graph isomorphism testing is useful for generation of molecular graphs and for computer synthesis. Chemical database 
search is an example of graphical data mining, where the graph canonization approach is often used. In particular, a number 
of identifiers for chemical substances, such as SMILES and InChI, designed to provide a standard and human-readable way to 
encode molecular information and to facilitate the search for such information in databases and on the web, use canonization 
step in their computation, which is essentially the canonization of the graph which represents the molecule.
In electronic design automation graph isomorphism is the basis of the Layout Versus Schematic (LVS) circuit design step, 
which is a verification whether the electric circuits represented by a circuit schematic and an integrated circuit layout 
are the same. Other application is the evolutionary graph theory, which is an area of research lying at the intersection of 
graph theory, probability theory, and mathematical biology. Evolutionary graph theory is an approach of studying how 
topology affects evolution of a population. That the underlying topology can substantially affect the results of the 
evolutionary process is seen most clearly in a paper by Erez Lieberman, Christoph Hauert and Martin Nowak.

So, it's important to design polynomial time algorithms to test if two graphs are isomorphic at least for some special classes of graphs. 
 An approach 
to this was  presented by Eugene M. Luks(1981) in the work \textit{Isomorphism of Graphs of Bounded Valence Can Be Tested in Polynomial 
Time}. Unfortunately,  it was a theoretical algorithm and was very difficult to put into practice.  
On the other hand, there is no known implementation of the algorithm, although Galil, Hoffman and Luks(1983) shows an improvement of 
this algorithm running  in $O(n^3 \log n)$. 

The two main goals of this  master thesis are to  explain more carefully the algorithm of Luks(1981), 
including a detailed study of  the complexity and, then to provide an efficient implementation in SAGE system.
It is divided into four chapters plus an appendix. 

Chapter 1 mainly presents the preliminaries needed to follow the rest of the dissertation. This chapter contains three sections, the first 
section introduces the topics about group theory, in particular  the symmetric group, and the second  one
introduces the main definitions and results  of graph theory. Then,  the last shows the 
complexity theory  concepts.

Chapter 2 is devoted to collect  some basic algorithms in group and graph theory for later use.

Chapter 3 is the main part, and it is dedicated to  clarify carefully the trivalent case and the complexity of the 
algorithm.  The last section  extends the algorithm to a general case. 

Finally, Chapter 4 deals with  the implementation test.  

Appendix A is dedicated to the documentation of the implementation in SAGE system.


\chapter{Preliminaries}
 
This chapter gives a gentle yet concise introduction to most of terminology used later in this master thesis.

\section{Group theory background} \nocite{rose1978course}

We will focus on the theory of groups concerning the symmetric group, for further background we refer the reader to \cite{rose1978course}.

The symmetric group\index{symmetric group} of a finite set A is the group whose elements are all the bijective maps from A to A and whose group operation is 
the composition of such maps. In finite sets, "permutations" and "bijective maps" act  likewise on the group, we call that action 
rearrangement of the elements. 

The symmetric group of degree n is the symmetric group on a set $A$, such as $|A|=n$, we will denote this group by $S_n$, or if the 
set $A$ requires explanation by $Sym(A)$. 

Since a cycle $ (i_1 \ldots i_r )$ can be written as a product of transpositions; $S_n$ is 
generated by its subset of transpositions. But, except for the case $n=2$, we don't need every transposition in order to generate the 
symmetric group, since for $1 \leq j< k < n$, we have
\[
 (j \; k+1) = (k \; k+1)(j \; k )(k \; k+1)
\]
 
Thus the transposition $(j\; k+1)$ can be obtained from $(j \; k)$ and $(k \; k+1)$. Therefore the subset 
\[
 S = \{ (i \; i+1 ) \; 1 \leq i < n \}
\]

consisting of the \emph{elementary transpositions}, generates $S_n$. A further system of generators of $S_n$ is obtained from the expression
\[
 ( 1 \ldots n)^i ( 1 \; 2 ) ( 1 \ldots n)^i = ( i+1 \; i+2) \qquad 1 \leq i \leq n-2
\]

so that we have proved that the symmetric group $S_n$ is generated by  permutations $(1\; 2)$ and $(1 \;\ldots \; n ) $.

 A \emph{permutation group}\index{permutation group} is a finite group $G$ whose elements are permutations of a given set and whose group operation is composition of permutations in $G$, i.e., a permutation group is a subgroup of the symmetric group on the given set.

We will say that a subset $T$ of $Sym(A)$ \emph{stabilizes}\index{stabilize} a subset $B$ of $A$ if $\sigma(B) = B$ for all $\sigma \in G$. If $G$ is a group and
$G$ stabilizes a subset $B$, we will say that $G$ acts on $B$, i.e. we have an homeomorphism from $G$ to $Sym(B)$. An action $G$ over $B$ 
is called \emph{faithful}\index{faithful} if the homomorphism is injective.


\begin{definition}\index{$G-$orbit}
If $G$ acts on $B$ and $b \in B$, the \emph{$G$-orbit} of $b$ is the set $G_b =\{ \sigma(b) \; | \; \sigma \in G \}$.
\end{definition}

We say that a group $G$ acts \emph{transitively} on $B$ if $B =G_b$, for some $b \in B$. Note that if $B=G_b$ for some $b \in B$, then $B= G_b$ for all 
$b \in B$.




\begin{definition}\index{$G$-block}
A \emph{$G$-block} is a subset $B$ of $A$,$B \neq \emptyset$, such that, for all  $\sigma \in G$, $\sigma(B) = B $ or $\sigma(B) 
 \cap B = \emptyset$.

\end{definition}

 In particular, the sets $A$ and all 1-element subsets of $A$ are blocks, these are called the trivial blocks. An example of  non-trivial blocks in 
a group that no act transitively on A, are the $G-$orbits \footnote{$\sigma(G_b) = G_b \forall \sigma \in G$}.

If $B$ is a $G-$block, then a \emph{$G-$block system}\index{$G-$block system} is the collection $\{ \sigma(B) \; | \; \sigma \in G \}$

\begin{example}
 Let $n=4$ and $G= \{ id, (13)(24),(14)(23),(12)(34) \}$ then the set $\{1,3\}$ is a $G-$block and the collection $\{ \{1,3\}, \{2,4\} \}$ is a
$G-$block system.
\end{example}

 The action $G$ is said to be \emph{primitive}\index{primitive} if the only $G-$blocks are the trivial blocks. We have that the $G-$orbits are 
$G-$blocks, so if $G \neq Id$ acts primitively on $A$ then $G$ acts transitively. In the case that $G$ acts transitively the $G-$blocks are 
called \emph{block of imprimitivity}\index{block of imprimitivity}.


A $G-$block system is said to be \emph{minimal}\index{minimal, $G-$block system} if $G$ acts primitively on the blocks. In the previous example 
the $G-$block system is minimal. Note that the number of blocks in a minimal $G-$block system is not, in general, uniquely determined. However, 
we have the next result.

\begin{lema}\label{MinimalBlockSystemLema}
 Let $P$ be a transitive $p-$subgroup of $Sym(A)$ with $|A|>1$. Then exists a $P-$block system consists of exactly $p$ blocks. 
Furthermore, the subgroup, $P'$, which stabilizes all of the blocks has index $p$.
\end{lema}

\begin{Proof}
 The quotient $P/P'$ is a primitive $p-$group (acting on the blocks) and so the order of $P/P' = $number of blocks $= p$ 
\cite[p.~66]{hall1976theory}
\qed  \end{Proof}

Thanks the above lemma,  if $P$ is a $2-$subgroup of $Sym (A)$, then exists $B_1, B_2$ such 
$A= B_1 \cup B_2$ where $B_1 $ and $B_2$ are $P-blocks$.

\section{Graph theory background}

Fortunately, much of standard graph theoretic terminology is so intuitive that it is easy to remember.

 A \emph{graph}\index{graph} is a pair $G=(V,E)$ of sets that $E \subseteq V^2$; thus, the elements of $E$ are $2-$element subsets 
of V. The elements of $V$ are the \emph{vertex} \index{vertex} of the graph $G$; and the elements of $E$ are the \emph{edges} \index{edge}.

\begin{note}
 If we consider vertices as $2-$tuples, we have a \emph{digraph} in the example below we can see the differences between a 
graph and a digraph.
\end{note}

\begin{example}
 Take $E=\{1,2,3,4 \}$ and $V= \{ (1,2) , (1,3) , (1,4) \} $ then the graph $G$ is the graph that we can see in \hyperlink{Grafexemple1}{Figure \ref{Grafexemple1}}
and the digraph is the graph that we can see in  \hyperlink{Digrafexemple1}{Figure \ref{Digrafexemple1}}

\begin{figure}[h!]
 \centering
\includegraphics[scale=0.6]{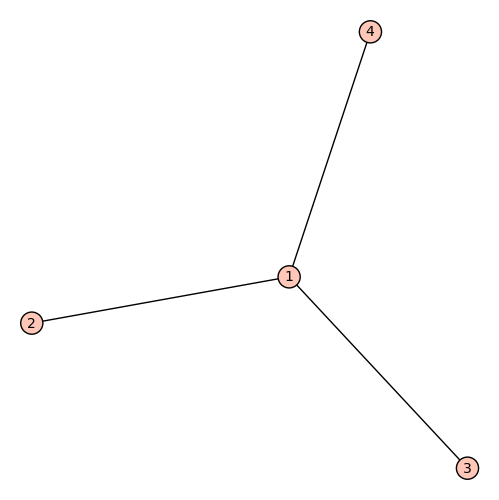}
\caption{Graph with $V=\{1,2,3,4 \}$ and $E=\{(1,2),(1,3),(1,4)\}$.}\label{Grafexemple1} \hypertarget{Grafexemple1}{}
\end{figure}

\begin{figure}[h!]
  \centering
\includegraphics[scale=0.6]{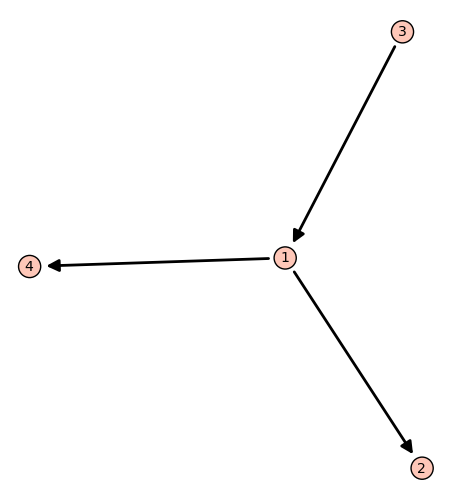}
\caption{Digraph with $V=\{1,2,3,4 \}$ and $E=\{(1,2),(3,1),(1,4)\}$.}\label{Digrafexemple1} \hypertarget{Digrafexemple1}{}
\end{figure}

\end{example}

\begin{note}
 Note that the graph $G=(\{ 1,2,3,4 \} , \{ (1,2), (1,3), (1,4) \}) $ is the same graph that $G' = ( \{ 1,2,3,4 \}, \{ (1,2), (3,1), (1,4) \} )$, 
but if we consider $G, G'$ as digraph they are not the same digraph.
\end{note}

The vertex set of a graph $G$ is referred to as $V(G)$ and the edge set as $E(G)$. These conventions are independent of any actual name 
of these two sets, for example if we define a graph $H=(W,F)$ the vertex set of the graph is still referred to as $V(H)$, not as $W(H)$.
If there is no possible confusion we don't distinguish between the graph and the vertex set or the edge set; for example we say a vertex 
$v \in G$ and an edge $e \in G$. 

\begin{definition}
If $G$ is a graph, then two vertices $e_1, e_2 \in E(G)$ are \emph{neighbors} if $(e_1, e_2) \in V(G)$. If we have a digraph we said that 
$e_1$ is a \emph{successor} of $e_2$ or, equivalently, $e_2$ is a \emph{predecessor} of $e_1$ if $(e_1, e_2) \in V(G)$ 
\end{definition}

Another well known concept is the following:

\begin{definition}\index{path}
 A \emph{path} in a graph is a sequence of vertices such that from each of its vertices there is an edge to the next vertex in the sequence. 
A \emph{cycle} \index{cycle} is a path such that the start vertex and end vertex are the same. The choice of the start vertex in a cycle is 
arbitrary.
\end{definition}

A special family of graphs are:
\begin{definition}\index{connected graph}
 In a graph $G$, two vertices $u$ and $v$ are called \emph{connected} if $G$ contains a path from $u$ to $v$. A graph is said to 
be \emph{connected} if every pair of vertices in the graph is connected. A directed graph is called \emph{weakly connected} if replacing all 
of its directed edges with undirected edges produces a connected graph.
\end{definition}

We also  need the following two concepts:
\begin{definition}\index{degree}
 In an undirected graph $G$, the \emph{degree} of a node $v \in V(G)$ is the number of edges that connect to it.
In a directed graph, the \emph{in-degree} of a node is the number of edges arriving at that node, and the \emph{out-degree} is the 
number of edges leaving that node.
\end{definition}

\begin{definition}\index{valence}
We define the \emph{valence} of an undirected graph $G$ as $\max_{v \in V(G)} ( deg(v))$
\end{definition}

Using the above definitions, we can state the following well known result: 
\begin{proposition}
Let $X$ a connected graph with valence $t$ then
\[|E(X_2)| < |V(X_2)| \cdot t \]
\end{proposition}

\begin{Proof}
Every node $v \in V(X)$ is connected with at most $t$ nodes, then for each node, are at most $t$ edges connected to $v$.
\qed  \end{Proof}

The following is a natural definition:

\begin{definition}\index{isomorphism}
 Let $G=(V,E)$ and $G'=(V',E')$ be two graphs or digraphs. We say that $G$ and $G'$ are \emph{isomorphic} if there exists a bijection 
$\varphi: V \rightarrow V'$ such as $(x,y) \in E \Leftrightarrow (\varphi(x), \varphi(y) \in E'$ for all $x,y \in V$.
\end{definition}

The previous map $\varphi$ is called an \emph{isomorphism}, if $G=G'$, it is called an \emph{automorphism}

\begin{proposition} \index{$Aut(G)$}
 Let $G=(V,E)$ a graph, the set of automorphisms, $Aut(G)$, define a permutations group.
\end{proposition}

\begin{Proof}
We only need see that $Aut(G)$ is a subgroup of $Sym(V)$.

\begin{itemize}
 \item  If $\varphi \in Aut(G)$ then $\varphi^{-1}$ is an automorphism because is bijective and if we have
\[
 (x,y) \in E \Leftrightarrow ( \varphi(x), \varphi(y) \in E 
\]
then if we apply $\varphi^{-1}$ in both edges

\[
 ( \varphi^{-1}(x), \varphi^{-1} (y) ) \in E \Leftrightarrow (x,y) \in E
\]
\item If $ \varphi, \varphi' \in Aut(G)$ then
\[
 (x,y) \in E \Leftrightarrow (\varphi(x), \varphi(y) ) \in E \Leftrightarrow ( \varphi' ( \varphi (x)), \varphi' ( \varphi (y))) \in E
\]

\end{itemize}
then $Aut(G)$ is closed under inverses and products, so $Aut(G)$ is a subgroup of $Sym(V)$ and therefore $Aut(G)$ is a permutation group.
\qed  \end{Proof}

The above result suggest the following notation.

\begin{definition}
We denote by $Aut_e(G)$ the subgroup of $Aut(G)$ such as fix the edge $e$, ie, $\forall \varphi \in Aut_e(G) $ if $e = (v_1, v_2)$ then 
$\varphi (v_1) = v_2$ and $\varphi(v_2) = v_1$ or $\varphi(v_1 ) = v_1$ and $\varphi (v_2) = v_2$.
\end{definition}

The following example illustrates the above concepts:
\begin{example}
 Let $G$ the graph of Example 1, then $Aut(G) = \langle (2,3), (2,4), (3,4) \rangle$ and if $e = (1,2)$, $Aut_e(G) = \langle (3,4) \rangle$.
If we consider the digraph, then $Aut(G) = \langle (2,4) \rangle$ and $Aut_e(G) = Id$. 
\end{example}

\begin{definition}
A \emph{tree}\index{tree} is a finite, connected, acyclic graph, we say that a tree is rooted if it has a distinguished node, called root. 
In a rooted tree, \index{parent}the \emph{parent} of a node $x$ is the unique node adjacent to $x$ which is closer to the root, \index{children} 
the \emph{children} of a node are the nodes of which $x$ is the parent; a node $x$ is an \emph{ancestor} \index{ancestor} of a node $y$ 
if the shortest path from $y$ to the root contains $x$, in this case we also say $y$ is a descendant of $x$. 
\end{definition}

In a tree $T$ we have two type of vertices: \emph{leaves} $L(T)$, terminal nodes, they belong to a single edge, in a rooted tree a \emph{leaf} 
is a node without children; and \emph{interior nodes} $Int(T)$

\begin{definition}\index{phylogenetic tree}
 A \emph{phylogenetic tree} is a triplet $(T, \rho, \{u_1, \ldots, u_n \} )$ where $T$ is a tree with $n$ leaves, $\{ u1, \ldots, u_n \}$ is a 
set of different species (or taxa), and $\rho: \{u_1, \ldots, u_n \} \rightarrow L(T)$ is a bijection.
\end{definition}

In the literature the leaves represent current species and the interior nodes represent ancestral species. The tree records the ancestral relationships 
among the current species.

\begin{definition} \index{evolutionary network}
By a \emph{evolutionary network} on a set $S$ of taxa we simply mean a rooted directed acyclic graph, with its leaves bijectively labeled 
in $S$. 
\end{definition}

A \emph{tree node}\index{tree node} of an evolutionary network $N=(V,E)$ is a node of in-degree at most $1$, and a \emph{hybrid node} is a
 node of in-degree at least 2. A \emph{tree arc} (\emph{hybridization arc} ) is a path such that the start vertex is a tree node (hybrid node). 
As in tree, a node $v\in V$ is a \emph{child} of $u\in V$ if $(u,v) \in E$, we also say in this case that $u$ is a \emph{parent} of $v$, note that 
in this case a node can have more than one parent.

\begin{definition}\index{binary evolutionary network}
 An evolutionary network is \emph{binary} when its hybrid nodes have in-degree 2, out-degree 1 and internal tree nodes have out-degree 2.
\end{definition}

An \emph{isomorphism} between two rooted trees $T_1$ and $T_2$ is an isomorphism from $T_1$ to $T_2$ as graphs that sends the root of $T_1$ to
the root of $T_2$. An isomorphism between phylogenetic trees or evolutionary networks also preserves the bijection $\rho$, ie, let $\varphi: V(T_1) \rightarrow V(T_2)$ an 
isomorphism between $(T_1, \rho_1, \{ u_1, \ldots, u_n \} )$ and $(T_2, \rho_2, \{ u_1, \ldots, u_n \} )$, then $\varphi(\rho_1(u_i))= \rho_2(u_i)$ 
$\forall i = 1, \ldots, n$. If $T_1$ and $T_2$ have roots $r1, r_2$ respectively, we also require that $\varphi(r_1) = r_2$.

\section{Computational complexity theory background}

Computational complexity theory is a branch of the theory of computation in theoretical computer science and mathematics that focuses on 
classifying computational problems according to their inherent difficulty, and relating those classes to each other. Many important complexity 
classes can be defined by bounding the time or space used by the algorithm. Some important complexity classes of decision problems defined by
bounding space are the following:

\begin{center}
\begin{tabular}[c]{c|c|c}
\textbf{Complexity class} & \textbf{Model of computation} & \textbf{Resource constraint}\\
\hline DTIME(f(n)) & Deterministic Tuning Machine & Time $f(n)$\\
P & Deterministic Tuning Machine & Time $poly(n)$\\
EXPTIME & Deterministic Tuning Machine & Time $2^{poly(n)}$\\
NTIME(f(n)) & Non-deterministic Tuning Machine & Time $f(n)$\\
NP & Non-deterministic Tuning Machine & Time $poly(n)$\\
NEXPTIME & Non-deterministic Tuning Machine & Time $2^{poly(n)}$
\end{tabular}
\end{center}

We will focus on the class P, also known as PTIME. PTIME is one of the most fundamental complexity classes, it contains all decision problems 
witch can be solved by a deterministic Turing machine using a polynomial amount of computation time, or polynomial time. Cobham's thesis holds 
that P is the class of computational problems which are ``efficiently solvable'' or ``tractable''; in practice, some problems not known to be in P have 
practical solutions, and some that are in P do not, nut this is a useful rule of thumb. 

A more formal definition of P is

\begin{definition}\index{P}
 A language $L$ is in P if and only if there exists a deterministic Turing machine $M$, such that
\begin{itemize}
 \item $M$ runs for polynomial time on all inputs
 \item For all $x \in L$, $M$ outputs 1
 \item For all $x \notin L$, $M$ outputs 0
\end{itemize}
\end{definition}

\subsection{Reducibility}
Intuitively, a problem $Q$ can be reduced to another problem $Q'$ if any instance of $Q$ can be ``easily rephrased'' as an instance of $Q'$, 
the solutions which provides a solution to the instance of $Q$. For example, the problem of solving equations linear equations in an 
indeterminate $x$ reduces to the problem of solving quadratic equations. Given an instance $ax+b=0$, we transform it to $0x^2 + ax+b=0$, 
whose solution provides a solution to $ax+b=0$.  Thus, if a problem $Q$ reduces to another problem $Q'$, then $Q$ is, in a sense, 
``no harder to solve'' than $Q'$.

\begin{definition}\index{Polynomial reduction}
If exists a polinomial-time algorithm $F$ that computes this ``rephrasing'', then we say that $Q$ is polynomial-time 
reducible to $Q'$.
\end{definition}

Then if we can solve the problem $Q'$ in polynomial time, we can solve the problem $Q$. This technique is very useful because, generally, is easy 
find a easier problem that is polynomial time reducible to our initial problem.

\begin{example}
Solving linear equations in an indeterminate $x$ clearly reduces in polinomial time to the problem of solving quadratic equations
\end{example}

\chapter{ Basic algorithms }

In this chapter we introduce some basic algorithms. The first section contains algorithms in group theory that we will use in the 
main algorithm. In the second section we will present an algorithm to test if two phylogenetic trees are isomorphic, with this example 
we will see that sometimes the isomorphism problem is easy.

\section{Algorithms in group theory}

Since every subgroup of $S_n$ can be generated by at most $n$ elements \cite{Mark198660}, any subgroup of $S_n$ can be specified in space 
which is polynomial in $n$.

\begin{lema}[Furst-Hopcroft-Luks]\label{LemaFilter}
 Given a set of generators for a subgroup $G$ of $S_n$ one can determine in polynomial-time
\begin{enumerate}
 \item the order of $G$
 \item whether a given permutation $\sigma$ is in $G$
 \item generators for any subgroup of $G$ which is known to have polynomially bounded index in $G$ and for which a polynomial-time membership
 test is available.
\end{enumerate}

\end{lema}

\begin{Proof}
 Let $G$ a subgroup of $S_n$, denote by $G_i$ the subgroup of $G$ which fixes the numbers in $\{1, \ldots, i \}$. Thus we have a chain of 
subgroups
\[
 1 = G_{n-1} \subseteq \cdots \subseteq G_1 \subseteq G_0 = G
\]

Now we construct a complete sets of coset representatives, $C_i = G_i $ modulo $G_{i+1}$ $0 \leq i \leq n-2$, then $|G| = \Pi_{i=0}^{n-2} |C_i |$. The 
main part of this construction is the subroutine \hyperlink{Filter}{Algorithm \ref{Filter}}. The input is an element $\alpha \in G$, 
the lists $C_i$ contain sets of left coset representatives for $G_i $ modulo $G_{i+1}$. 

Thus the subroutine searches for a representative of the coset of $\alpha$ modulo $G_{i+1}$ in the list $C_i$. If it is not found, then $\alpha$ 
represents a previously undiscovered coset and it is added to the list. If it is found as $\gamma$ then $\gamma^{-1} \alpha$ is in $G_i$ and its 
class modulo $G_{i+1}$ is sought in $C_i$. Since, for $\sigma \in G_i$, membership in $G_{i+1}$ is testable in constant time ( we only need see if 
$\sigma ( i+1 ) = i+1$ ), the procedure requires only polynomial time.

The algorithm for the first part of lemma is now easily stated:
\begin{enumerate}
 \item Initialize $C_i \leftarrow \{ 1 \}$ for all $i$.
 \item Filter the set of generators of $G$.
 \item Filter the sets $C_i C_j$ with $i \geq j$.
\end{enumerate}

\begin{algorithm}\label{Filter}\hypertarget{Filter}{}
 
\KwData{$\alpha \in G$}
\KwResult{Add $\alpha$ to his $C_i$}
\Begin{
\For{$i \in [0,n-2]$}{
    \If{$\exists \gamma \in C_i \mbox{ : }\gamma^{-1} \alpha \in G_{i+1} $}{
	  $\alpha \leftarrow \gamma^{-1} \alpha $
	}
     \Else{ add $\alpha$ to $C_i$ \\
	    \Return{}
      }
}

\Return{}
}
\caption{Filter} \label{Filter}\hypertarget{Filter}{}

\end{algorithm}

Of course, the calls to the subroutine may result in an increase in some $C_i$, thus demanding more runs of (3). However, we know a priori that,
at any stage, $ |C_i | \leq | G_i : G_{i+1} | \leq n-i$. Thus the process terminates in polynomial time. The result of (2) is that the 
original generating set is contained in $C_0C_1 \ldots C_{n-2}$. The actual outcome of (3), given (1), is that $C_i C_j \subseteq C_jC_{j+1}
\cdots C_{n-2}$. These facts can be used to prove that $G=C_0C_1\ldots C_{n-2}$. That $C_i$ represents $G_i$ modulo $G_{i+1}$ is then immediate.

By the first part of lemma, the second is an immediate consequence of the fact: $\sigma \in \langle \Phi \rangle \Leftrightarrow 
 |\langle \Phi, \sigma \rangle | = | \langle \Phi \rangle |$. Wehave  that this membership test might be implemented by a construction of the 
lists $C_i$ for $\langle \Phi \rangle$ followed by the call Filter($\sigma$). Then $\sigma \in G $ if and only if it doesn't force an increase in 
some $C_i$.

For the last part of lemma, we alter the group chain to
\[
 1 = H_{n-1} \subseteq \cdots H_2 \subseteq H_1 \subseteq H \subseteq G
\]

and apply the same algorithm to generate complete sets of coset representatives. Note that the polynomial index of $H$ in $G$ and the requirement 
that the membership in $H$ be polynomially decidable guarantees again that the entire process takes only polynomial time. Ignoring the first list, 
the remaining lists comprise a set of generators for $H$.
\qed  \end{Proof}

\begin{remark}
 The complexity of the algorithm \texttt{Filter} is $O(n^5)$ because at most there are $O(n^4)$\footnote{$\sum_{0 \leq i \leq j \leq n-2} (n-i)(n-j)$ is in $O(n^4)$} 
 elements in the union of $C_i$ and  we need an extra $n$ to check whether an element is in its corresponding $C_i$. The complexity 
 of the third part of the lemma is $O(n^5) \cdot O( \mbox{test membership in } H)$
\end{remark}

We will need, in the transitive case, to be able to decompose the set into non-trivial blocks of imprimitivity. To be precise, we fix $a \in A$ and 
for each $b \in A$, $b \neq a$, we generate the smallest $G-$block containing $\{a,b \}$.

\begin{proposition}  (\cite{Sims1967})
 The smallest $G-$block containing $\{a,b\}$ is the connected component of $a$ in the graph $X$ with $V(X)=A$ and $E(X)$ is the $G-$orbit of 
$\{a,b\}$ in the set of all (unordered) pairs of elements of $A$.
\end{proposition}

If $G$ is imprimitive, the block must be proper for some choice of $b$, in that case, the connected components of $X$ define a $G-$block system.
Repeating the process, we actually obtain an algorithm for the following computational problem.

\begin{lema}
 Given a set of generators for a subgroup $G$ of $S_n$ and a $G-$orbit $B$, one can determine in polynomial time, a minimal $G-$block system in $B$.
\end{lema}

Thanks to Atkinson \cite{Atkinson1975}, we have the \hyperlink{Blocks}{Algorithm \ref{Blocks}}, that is a particularly efficient implementation of 
the above ideas. 

\begin{algorithm}\label{Blocks}\hypertarget{blocks}{}
\KwData{$\omega \neq 1$, $G = \langle g_1, \ldots, g_m \rangle$}
\KwResult{The smallest $G-$block which contains $\{1, \omega \}$}
\Begin{
$C \leftarrow \emptyset$ \\
Set $f( \alpha = \alpha) \; \forall \alpha \in A$ \\
Add $\omega$ to $C$ \\
Set $f(\omega) = 1$ \\
\While{$C$ is nonempty}{
Delete $\beta$ from $C$ \\
$\alpha \leftarrow f( \beta) $\\
$j \leftarrow 0 $ \\
\While{$j < m$}{
$j++$ \\
$\gamma \leftarrow \alpha g_j$ \\
$\delta = \beta g_j$ \\
\If{$f(\gamma) \neq f( \delta)$}{
Ensure $f( \delta) < f(\gamma)$ by interchanging $\gamma$ and $\delta$ if necessary. \\
\For{ $\epsilon \; : \; f( \epsilon ) = f( \gamma )$}{
      Set $f( \epsilon) = f( \delta )$
}
Add $f( \gamma)$ to $C$.
}
}
}
\Return{C}
}
\caption{Smallest $G-$block which contains $\{1, \omega \}$}
\end{algorithm}

Let $f_0$ be the initial function $f$, and $f_1, \ldots, f_r = \bar{f}$ be the variants of $f$ defined by the last \texttt{For}. Associated with 
each function $f_i$ is a partition $P_i$ of $A$, each part of $P_i$ consists of elements on which $f_i$ takes the same value, ie, if $B \in P_i$, 
then $\forall \alpha, \beta \in B$, $f_i (\alpha ) = f_i ( \beta)$.  Also we have that $P_{i+1}$ is obtained from $P_i$ by replacing to parts of $P_i$
 by their union; in particular, $P_i$ is a refinement of $P_{i+1}$ as every part of $P_i$ is contained in a part of $P_{i+1}$.

We denote by  $P_i ( \alpha)$  the part of $P_i$ which contains $\alpha$.

\begin{lema}
 \begin{enumerate}
  \item If $f_i( \alpha ) = f_i ( \beta) $ then $f_j ( \alpha ) = f_j ( \beta)$, $\forall j \geq i$.
 \item $f_i ( f_i ( \alpha ) ) = f_i ( \alpha)$ $\forall \alpha \in A $ and $\forall i \geq 0$.
 \end{enumerate}

\end{lema}
\begin{Proof}
 \begin{enumerate}
  \item The proof of this part is obvious by construction, because if we change $f_j( \alpha)$ we also change $f_j( \beta)$ by the same value.
  \item Clearly $f_0 ( \alpha )  P_0 ( \alpha)$ for all $\alpha$, and it is also evident that $f_i ( \alpha ) \in P_i( \alpha)$, so $ \alpha , 
f_i ( \alpha) \in P_i ( \alpha)$ then  $f_i ( \alpha ) = f_i ( f_i ( \alpha))$.
 \end{enumerate}
\qed  \end{Proof}

\begin{lema}
 \begin{enumerate}
  \item $\alpha \geq f_0 ( \alpha) \geq f_1 ( \alpha) \geq \cdots \geq \bar{f} ( \alpha)$
  \item A point $\beta$ belonged to $C$ if and only if $\beta \neq \bar{f} ( \beta)$.
  \item If $ \beta$ belonged to $C$, then there exists $\alpha < \beta $ with $\bar{f} ( \alpha ) = \bar{f} ( \beta)$ and $\bar{f} ( \alpha g_j ) = 
        \bar{f} ( \beta g_j )  \; ,  \; j = 1, \ldots , m$.
 \end{enumerate}

\end{lema}

\begin{Proof}
 \begin{enumerate}
  \item The first step ensures that $\alpha \geq f_0 ( \alpha)$ and the step before the \texttt{For} instruction ensures that $f_i ( \alpha ) \geq 
        f_{i+1} ( \alpha )$.
 \item  The points of $C$ are added in the line 4 and 18. In the line 4, $\beta = \omega$ and $ \omega  > f_0 ( \omega) = 1 = \bar{f}( \omega)$. 
In the line 18, $\beta = f_i ( \gamma)$ for some $i $ and $ \gamma$; then $f_i ( \beta) = \beta = f_i ( \gamma)$ and $f_{i+1} ( \beta) < f_i (\beta)$. 
Conversely, if $\beta > \bar{f} ( \beta)$; then clearly $f_i ( \beta ) = \beta $ belonged to $C$.

\item Let $\alpha$ be the point defined in line 8, when $\beta$ is deleted from $C$. Then $\alpha = f_i ( \beta) < \beta $ for some $i$. 
Moreover, by the previous lemma, $f_i ( \alpha) = f_i^2( \beta)$ and so $\bar{f}(\alpha) = \bar{f}(\beta)$. Finally, after line 18 for a given 
$j$, $f_k ( \alpha g_j ) = f_k ( \beta g_j)$ for some $k$ and so $ \bar{f} ( \alpha g_j ) = \bar{f} ( \beta g_j )$. 
 \end{enumerate}

\qed  \end{Proof}

\begin{lema}
 $\bar{P} = P_r$ is invariant under $G$.
\end{lema}
\begin{Proof}
 It is sufficient to prove that each $g_j$ preserves $\bar{P}$. Suppose that there exists $a,b \in A$, $a \neq b$ with $\bar{f}(a) = \bar{f} (b)$ but $\bar{f} ( g_j (a)) 
\neq \bar{f}(g_j(b))$,  with  $b$  minimal. Then $\bar{f}(b) = \bar{f}(a) \leq a < b$ and so $b$ belonged to $C$. Hence there exists $c < b$ with 
$\bar{f}(c) = \bar{f}(b)$ and $\bar{f}( g_j(c)) = \bar{f}(g_j (b))$. Since $\bar{f}( c ) = \bar{f}(a)$ and $c < b$, ensures that $\bar{f}( g_j ( c) ) = \bar{f}(g_j(a))$.
Thus $\bar{f}( g_j ( b) ) = \bar{f}(g_j(c)) = \bar{f}(g_j(a))$ it is a contradiction.
\qed  \end{Proof}

So $\Delta = \bar{P}(1)$ is a block of $G$ containing $1$ and $\omega$.  As $G$ is transitive and the previous lemma states  that $\bar{P}$ is 
$G-$invariant, then $\bar{P}$ is the block system containing $\Delta$.

\begin{lema}
 $\Delta$ is the smallest block containing $1$ and $\omega$.
\end{lema}

\begin{Proof}
 Let $\Delta_1$ be the smallest block containing $1$ and $\omega$ such that $\Delta_1 \subseteq \Delta$. Let $\hat{P} = \{ g( \Delta_1 ) \; | \; g \in G \}$. 
Then $\hat{P}$ is a partition of $A$; we now prove that each $P_i$ is a refinement of $\hat{P}$ by induction on $i$. 
This is clearly true if $i = 0$. Assume now that $i>0$, $P_i$ is a refinement of $\hat{P}$ and consider a part of $P_{i+1}$. 
Such a part is either a part of $P_i$ or the union of 
two parts of $P_i$ of the form $P_i ( f_i ( \ gamma)) \cup P_i ( f_i ( \delta)) = P_i ( \gamma) \cup P_i ( \delta)$ where $\gamma = \alpha g_j , \delta = \beta g_j$ 
and $P_i ( \alpha ) = P_i ( \beta)$. By an inductive assumption, $\hat{P} ( \alpha ) = \hat{P} ( \beta)$. Then 
\[
 \hat{P} ( \gamma) = \hat{P}( g_j ( \alpha)) = \hat{P} ( g_j ( \beta)) = \hat{P} ( \delta) \supseteq P_i ( f_i ( \gamma)) \cup P_i ( f_i ( \delta ))
\]

This completes the induction and we have $\Delta = \bar{P} (1) = P_r ( 1) \subseteq \hat{P} (1) = \Delta_1$. 
\qed  \end{Proof}

\begin{observation}
 There are several ways in which the algorithm can be made faster, we can see it in \cite{Atkinson1975}.
\end{observation}

In our applications it will be necessary to determine the subgroup of $G$ which stabilizes all of the blocks.

\begin{lema}
 Given a set of generators for a subgroup $G$ of $S_n$ and a $G-$orbit $B$, one can determine, in polynomial time, a set of generators for the 
subgroup of $G$ which stabilizes all of the blocks in a $G-$block system in $B$.
\end{lema}
\begin{Proof}
 The third part of the lemma \ref{LemaFilter} guarantees this. Let $G_i$ denote the subgroup which stabilizes each of the first $i$ blocks. Then ( taking $G= G_0$ ) 
\[
 | G_i : G_{i+1} | \leq \mbox{ number of blocks } - i 
\]

\qed  \end{Proof}

\section{Algorithms in graph theory}

If two rooted phylogenetic trees are isomorphic can be tested easily, using the extra information that we have ( $\varphi ( \rho_1 (u_i))= \rho_2 (u_i)$.
Thanks to this extra information we have the \hyperlink{PhyTreeIso}{Algorithm \ref{PhyTreeIso}}

\begin{algorithm}\label{PhyTreeIso}\hypertarget{PhyTreeIso}{}
 
\KwData{$T_1= (T_1, \rho_1, \{u_1, \ldots, u_2 \}) and T_2=(T_2, \rho_2, \{u_1, \ldots, u_2 \})$}
\KwResult{Test if $T_1$ and $T_2$ are isomorphic}
\SetKwFunction{PostOrderIterator}{PostOrderIterator}
\SetKwFunction{parent}{parent}
\Begin{
Set $\varphi ( \rho_1(u_i)) = \rho_2(u_i) \; \forall i$ \\
Nodes $\leftarrow$ \PostOrderIterator{$T_1$} \\
$w \leftarrow Nodes.next()$ \\
\While{Nodes.hasNext()}{
    \If{ $w$ is not a leaf}{
	\If{ $\varphi (w) == $none}{
	      $v$ child of $w$ \\
	      Set $\varphi(w) = parent( \varphi(v ) )$ 
	    }
	  \For{ $v$ child of $w$}{
	      \If{ $ \varphi(w) \neq $\parent{$\varphi (v) $}}{
		  \Return{False}
		  }
	      }
      }
}

\Return{$\varphi$}
}
\caption{PhylogeneticTreeIsomorphism} 

\end{algorithm}

The subroutine \texttt{PostOrderIterator} returns an iterator of the nodes of $T_1$ in postorder, i.e., first the leaves, then the parents of the leaves
and so to get to the root.

\begin{lema}
 The previous algorithm terminates in linear time.
\end{lema}
\begin{Proof}
 Let $n$ the number of leaves and $m = | V(T_1)|$, then in the algorithm we first made the iterator \texttt{PostOrderIterator} it can be made in 
$O(m)$, because each node has to be visited at least once and increases linearly for increasing $m$. Then in the loop, first we do $n$ trivial operations, 
corresponding to assigning $v \in L(T_1)$ to its corresponding $v' \in L(T_2)$, this operation is $O(n)$. Then for every $w \in Int(T_1)$ we 
do $O( child(w))$ operations to check if the isomorphism is correct, so we made $O(k)$ operations, were $k =\displaystyle\sum_{w \in Int(L)} 
| child(w)| = (m-n) \frac{ m-1}{m-n} = m-1$. Then the complexity is $O(n) + O(n) + O(m-1) = O(m)$.
\qed  \end{Proof}

This algorithm can be used to test if two evolutionary network are isomorphic, because we can reduce the size of the network by removing the part 
that is tree-like.

\chapter{Trivalent Case}

In this chapter we will see an extend explication of the problem when the valence of the graphs is 3, and at the end of chapter 
we will show a generalization to general case. The cases with $n=1$ and $n=2$ are trivial because for $n=1$ we only have one connected 
graph with valence 1, the graph with 2 nodes linked by 1 edge; and the case $n=2$ we only have two types of connected graphs, the 
``triangle'' with 3 nodes and 3 edges, and the list with $n$ nodes and $n-1$ edges.

\section{Reduction to the Color Automorphism Problem}
 
 We start reducing this  graph problem to a group one.
 
\begin{proposition}
 Testing isomorphism of graphs with bounded valence is polynomial-time reducible to the problem of determining generators for $Aut_e(X)$, where 
$X$ is a connected graph with the same valence, and $e$ is a distinguished edge.
\end{proposition}

\begin{Proof}
 First, we show  that if we can obtain a set of generators of $Aut_e(X)$ then we can test if two connected graphs of bounded valence  are isomorphic.
Let $e_1  \in E( X_1)$, then for $e_2 \in E(X_2)$  we can test if it exists an isomorphism from $X_1$ to $X_2$ sending  $e_1$ to $e_2$, as we can see in 
\hyperlink{IsotoAute}{Algorithm \ref{IsotoAute}}. We build the new graph from the disjoint union $X_1 \cup X_2$ as follows:
\begin{enumerate}
 \item Insert new nodes $v_1$ in  $e_1$ and $v_2$ in $e_2$.
 \item Join $v_1$ to $v_2$ with a new edge $e$. 
\end{enumerate}

\begin{algorithm}\label{IsotoAute}\hypertarget{IsotoAute}{}
\KwData{$X_1, X_2$ connected graphs of bounded valence}
\KwResult{Test if $X_1$ and $X_2$ are isomorphic}
\SetKwFunction{BuildX}{BuildX}
\SetKwFunction{Aut}{Aut}
\Begin{
$e_1 \in \mathcal{E}(X_1)$ \\
\For{$e_2 \in \mathcal{E}(X_2)$}{
	$X \leftarrow $ \BuildX{$X_1, X_2, e_1, e_2 $}\\
	$G \leftarrow $ \Aut($X, e $) \\
	\For{ $\sigma \in G $}{
	    \If{ $\sigma(v_1) == v_2 $}{
		\Return{ True}
	    }
	}
    }

\Return{False}
}
\caption{Isomorphism of graphs of bounded valence}

\end{algorithm}

\qed  \end{Proof}

\begin{obs}
 The \hyperlink{IsotoAute}{Algorithm \ref{IsotoAute}} works because if such automorphism  does exist, then any set of generators of $Aut_e(X)$ will contain one.
\end{obs}

Let $X_1$ and $X_2$ two connected trivalent graphs with $\frac{n-2}{2}$ vertices and build $X$ as before, then $X$ is a connected trivalent graph 
with $n$ vertices. The group $Aut_e(X)$ is determined through a natural sequence of successive ``approximations'', $Aut_e(X_r)$ where $X_r$ is the 
subgraph consisting of all vertices and all edges of $X$ which appear in paths of length $\leq r$ through $e$, more formally, if $e= (a,b)$

\[V(X_1) = \{ a,b \} \; , \; E(X_1) = \{ (a,b) \}\]
\[ V ( X_r) = \{ b \in V( X) \; | \; \exists a \in V(X_{r-1} ) \mbox{ such that} (a,b) \in E(X) \} \]
\[
 E(X_r) =  \{ ( a,b) \in E( X) \; | \; \exists a \in V(X_{r-1} ) \mbox{ such that} (a,b) \in E(X) \}
\]

There are natural homomorphisms
\[
 \pi_r : Aut_e(X_{r+1}) \rightarrow Aut_e(X_r)
\]
in which $\pi_r(\sigma)$ is the restriction of $\sigma$ to $X_r$. Now we construct a generating set for $Aut_e(X_{r+1})$ given one for $Aut_e(X_r)$.

For this we will solve two problems:
\begin{enumerate}[(I)]
 \item Find a set $\mathcal{K}$, of generators for $K_r$, the kernel of $\pi_r$. 
 \item Find a set $\mathcal{S}$, of generators for $\pi_r(Aut_e(X_{r+1}))$, the image of $\pi_r$.
\end{enumerate}

So, the algorithm to compute $Aut_e(X)$ is:

\begin{center}
 
\begin{algorithm}\label{Aute}\hypertarget{Aute}{}
\KwData{A sequence of graphs $Y$, whose are the result of \texttt{BuildX}}
\KwResult{$Aut_e(X)$ where $X$ is the last graph in the sequence}
\SetKwFunction{Ker}{Ker}
\SetKwFunction{Image}{Image}
\SetKwFunction{Pullback}{Pullback}
\Begin{
$Aut_e = ( e_1 \; e_2 )$
\For{$X \in Y$}{
	$K \leftarrow $ \Ker($X$)\\
	$S \leftarrow $ \Image($Aut_e, X$) \\
	$S2 \leftarrow $ \Pullback($S,X$)\\
	$Aut_e = S2 \cup K$
    }
\Return{$Aut_e$}
}
\caption{The group $Aut_e$}

\end{algorithm}

\end{center}

Then, if $\mathcal{S}'$ is any pullback of $\mathcal{S}$ in $Aut_e(X_{r+1})$, i.e. $\pi_r(\mathcal{S}') = \mathcal{S}$, then  $\mathcal{K} \cup \mathcal{S}'$ generates
 $Aut_e(X_{r+1})$.

Set $V_r = V(X_r) \setminus V(X_{r-1})$. Each vertex in this set is connected to one, two or three vertices in $X_r$. We codify this relationships 
as follows: Let $A_r$ denote the collection of all subsets of $V_r$ of size one, two, or three. Define
\[
 f: V_{r+1} \rightarrow A_r
\]
by $f(v) \{ w \in V(X_r) \; | \; (v,w) \in E(X) \}$, ie the \emph{neighbor set} of $v$.
 
\begin{definition}\index{twins}
 A pair $u,v \in V_{r+1}, u \neq v$, will be called \emph{twins} if they have the same neighbor set
\end{definition}

\begin{observation}
 There cannot be three distinct vertices with common neighbor set, because $X$ is a trivalent graph.
\end{observation}

\begin{proposition}
 \[
  \sigma \in Aut_e(X_{r+1}) \Rightarrow f( \sigma (v)) = \sigma ( f(v))
 \]

\end{proposition}

\begin{Proof}
Let $\sigma \in Aut_e(X_{r+1})$, then $\sigma$ preserves the set of edges so,
\[
w \in f(v) \Leftrightarrow ( w,v) \in E(X_{r+1}) \Leftrightarrow ( \sigma(w) , \sigma(v) ) \in E(X_{r+1}) 
\Leftrightarrow \sigma(w) \in f( \sigma (v))
\]
therefore $f(\sigma(v) ) = \sigma(f(v))$.
\qed  \end{Proof}  

In particular, if $\sigma \in \ker( \pi_r)$, $\sigma ( f(v)) = f(v)$, then $f(v) = f( \sigma (v))$, so either $v = \sigma (v)$ or $v$ 
and $\sigma(v)$ are twins. Since a permutation in $\ker( \pi_r)$ fixes neighbors sets of all $v \in V_{r+1}$, its only nontrivial action 
can involve switching twins. For each pair, $u,v$ of twins in $V_{r+1}$, let $(u \; v ) \in Sym (V( X_{r+1}))$ be the transposition that 
switches $u$ and $v$ while it fixes all other points. Problem (I) is solved by taking $\{ (u \; v ) \; | \; \mbox{ such that } u \mbox{ and } 
v \mbox{ are twins } \}$ for $\mathcal{K}$.

\begin{proposition}[Tutte]\label{Tutte}
 For each $r$, $Aut_e(X_r)$ is a $2-$group.
\end{proposition}

\begin{Proof}
 Since $| Aut_e(X_{r+1}) | = | Im \; \pi_r | \cdot | K_r |$, $K_r$ is the elementary abelian $2-$group generated by the transpositions in 
each pair of twins and a subgroup of $2-$group is a $2-$group; an induction argument recovers.
\qed  \end{Proof}

We note that if $\sigma \in Aut_e(X_r)$ is in $\pi_r(Aut_e(X_{r+1}))$, then it stabilizes each of the following 
three collections:
\begin{enumerate}
 \item The collection of edges ( considered as unordered pairs of vertices) connecting vertices in $V_r$:
\[
 A' = \{ (v_1, v_2, ) \in A \; | \; ( v_1, v_2 ) \in  E(X_{r+1}) \}
\]

 \item The collection of subsets of $V_r$ that are neighbor sets of exactly one vertex in $V_{r+1}$.
\[
 A_1 = \{ a \in A \; | \; a = f(v) \mbox{ for some unique } v \in V_{r+1} \}
\]
 \item The collection of subsets of $V_r$ that are neighbor sets of exactly two vertices in $V_{r+1}$, ie, the ``fathers'' of twins:
\[
 A_2 = \{ a \in A \; | \; a=f(v_1) = f(v_2 ) \mbox{ for some } v_1  \neq v_2 \}
\]

\end{enumerate}

Even more, this condition characterizes the set $\pi_r (Aut_e (X_{r+1}))$.

\begin{proposition}
$\pi_r(Aut_e(X_{r+1}))$ is precisely the set of those $\sigma \in Aut_e (X_r)$ which stabilize each of the collections $A_1, A_2, A'$.
\end{proposition}

\begin{Proof}
 We need only show that, if $\sigma$ stabilizes $A_1, A_2,A'$  then it does indeed extend to an element of $Aut_e(X_{r+1})$. For such $\sigma$, 
we define the extension as follows. For each ``only child'' $v,\; f(v) \in A_1$ we have $\sigma ( f(V)) \in A_1$, so we send  $v$ to the ``only child'' $v'$ 
such that $f(v') = \sigma(f(v))$. For each pair of twins $v_1, v_2$, $f(v) \in A_2$ implies $ \sigma(f(v)) \in A_2$, so map $\{ v_1, v_2 \}$ to the 
twins sons of $\sigma( f(v_1)) = \sigma(f(v_2))$ in either order. By construction, this extension stabilizes the set of edges between $V(X_r)$ and 
$V_{r+1}$. Note that $|f(v)|= |\sigma ( f(v))|$ also stabilizes the edges between ``old points'', because $\sigma$ stabilizes the set $A'$.
\qed  \end{Proof}

\begin{remark}
 We can not apply the {\sc Filter algorithm}, because we have no guarantee that the index 
  of the group that stabilizes the sets $A_1, A_2$ and $A'$ has a polynomial bound.
  \end{remark}

Now, set $B_r = V(X_{r-1}) \cup A_r$ and $G_r = Aut_e(X_r)$ and extend the action of $G_r$ to $B_r$, ie, if $v \in B_r$, $\sigma(v) = \{ \sigma(w) \; | \; w \in B_r \}$. 
To find $\mathcal{S}$, we color each element of $B_r$ with one of five colors that distinguish:
\begin{enumerate}[i)]
 \item whether or not it is in $A'$
 \item whether it is in $A_1$, or $A_2$ or neither.
\end{enumerate}
Only five colors are needed, since collections $A'$ and $A_2$ are disjoint when $r>1$, ie, let $C'  = B_r \setminus A', C_1 = B_r \setminus A_1$ 
and $C_2 = B_r \setminus A_2$, then the colors are:
\begin{enumerate}
 \item $A' \cap A_1 $ 
 \item $A' \cap C_1 $
 \item $C' \cap A_1 $
 \item $C' \cap A_2 $
 \item $C' \cap C_1 \cap C_2$
\end{enumerate}

We have $\sigma \in \pi_r ( Aut_e(X_{r+1}))$ if and only if $\sigma$ preserves colors in $A_r$. Thus, Trivalent Graph Isomorphism  problem is polynomial-time 
reducible to the following:

\begin{problem}\label{Problem3}
 \texttt{Input}: A set of generators for a $2-$subgroup $G$ of $Sym(A)$, where $A$ is a colored set. 

 \texttt{Find}: A set of generators for the subgroup $\{ \sigma \in G \; | \; \sigma \mbox{ is color preserving } \}$.
\end{problem}

\section{The Color Automorphism Algorithm for 2-Groups}

With a view toward a recursive divide-and-conquer strategy, we generalize the Problem \ref{Problem3}:

\begin{problem}\label{Problem4}
 \texttt{Input}: Generator for a 2-subgroup $G$ of $Sym(A)$, a $G-$stable subset $B$, and $\sigma \in Sym(A)$
 \texttt{Find}: $C_B( \sigma G)$.
\end{problem}
where $C_B(T) = \{ \sigma \in K \; | \; \sigma \mbox{ preserves the color } \forall b \in B \} $  
Problem \ref{Problem3} is an instance, with $B=A$, $\sigma=id$, of the Problem \ref{Problem4}.

Let $T, T'$ subsets of $Sym(A)$, and $B, B'$ subsets of $A$, then we have:
\begin{itemize}
 \item $C_B( T \cup T' ) = C_B (T) \cup C_B(T')$
 \item $C_{B \cup B' } (T) = C_B ( C_{B'} (K))$
\end{itemize}

We observe first that if $G$ is a subgroup of $Sym(A)$, and $B$ is a $G-$stable subset, then $C_B(G)$ is a subgroup of $G$. Also, we have the following lemma, needed  for  the recursive algorithm.

\begin{lema}\label{LeftCoset}
 Let $G$ be a subgroup of $Sym(A)$, $\sigma \in Sym(A)$ and $B$ a $G-$stable subset of $A$ such that $C_B( \sigma G)$ is not empty,  then it is a left coset of the subgroup $C_B(G)$.
\end{lema}

\begin{Proof}
 If $\sigma' \in C_B(\sigma G)$, then $\sigma G = \sigma' G$, because $\sigma' \in \sigma G$.  For $\tau \in G, b \in B$ we have that $\sigma' ( \tau (b))$ 
 has the 
same color as $\tau (b)$, because  $\tau (b) \in B$. Thus $\sigma' \tau \in C_B ( \sigma' G )$ if and only if $\tau \in C_B(G)$. That is,
$C_b( \sigma' G) = \sigma' C_B( G)$
\qed  \end{Proof}

Thanks to the lemma \ref{MinimalBlockSystemLema},\ref{LemaFilter} and \ref{LeftCoset}, we can present an algorithm for the problem \ref{Problem4}.

\begin{algorithm} \label{CB(G)}\hypertarget{CB(G)}{}
\KwData{ Coset $\sigma G \subseteq Sym(A)$ where $A$ is a colored set and $G$ a $2-$group, and a $G-$stable subset, $B$, of $A$.}
 \KwResult{ $C_B( \sigma G)$}
 
\Begin{
\Case{$B = \{ b \} $}{\eIf{ $\sigma(b) \sim b$}{ $C_B ( \sigma G ) = \sigma G$}{ $C_B = \emptyset $}
     }
\Case{ $G$ is intransitive on $B$}{
    Let $B_1$ a nontrivial orbit \\
    $B_2 = B \setminus B_1$ \\
    $C_B ( \sigma G ) = C_{B_2} ( C_{B_1} ( \sigma G)) $ 
   }
\Case{ $G$ is transitive on $B$}{
      Let $\{B_1, B_2 \}$ a minimal $G-$block system \\
      Find the subgroup , $H$, of $G$ that stabilizes $B_1$ \\
      Let $\tau \in G \setminus H$ \\
      $C_B ( \sigma G) = C_{B_2} ( C_{B_1}  (\sigma H)) \cup C_{B_2} ( C_{B_1}  (\sigma \tau H))$ \\
       }

\Return{ $C_B( \sigma G)$}
}
\caption{$C_B (\sigma G ) $}
\end{algorithm}

Observe that in the case  $G$ is transitive on $B$, we don't need to calculate $C_{B}( \sigma H)$ and $C_B ( \sigma \tau H )$, so,
 thanks to lemma \ref{LeftCoset}, we know, when $C_B( \sigma H)$ and $C_B( \sigma \tau H)$ both are non-empty sets, that  exists $\rho_1$ and 
$\rho_2 $ such as
\[
 C_B( \sigma H) = \rho_1 \sigma_B(H) \qquad C_B ( \sigma \tau H)= \rho_2 C_B( H)
\]

Then, form a generating set for $C_B(G)$ by adding $\rho_1^{-1}$ to the generators of $C_B(H)$, and take $\rho_1$ as the coset representative 
for $C_B( \sigma G)$.

\begin{proposition}
 The previous algorithm runs in polynomial time.
\end{proposition}

\begin{Proof}
 It is an standard induction argument.
 \qed  \end{Proof}

\begin{remark}
 In the next section we will see an upper bound of cost of this algorithm.
\end{remark}

\section{Study of complexity}

In this section we will prove the trivalent graphs  isomorphic problem is  polynomial time,  and we obtain an upper bound 
for the complexity using the Algorithm \ref{IsotoAute}. In order to do this we will divide the algorithm into parts, first we will 
compute the complexity of the algorithm \texttt{BuildX}, then  the complexity of the algorithm \texttt{Aut}, which we will separate 
in the algorithm \texttt{Ker},  \texttt{Image} and \texttt{Pullback}. Finally,  we will add a ``exponent'' because we will do this 
for all $e_2 \in \mathcal{E}(X_2)$ in the worst case, and we have a $O(3n)=O(n)$ edges in $X_2$, because the valence of $X_2$ is 3 
and we have at most 3 edges for every node.

\subsection{Algorithm \texttt{BuildX}}

In this algorithm we will build a sequence of graphs and the cost of the algorithm is the cost of building this sequence. We will 
assume that the cost of know the neighbors of a vertex is $O(1)$, assuming that the cost of building the sequence is $O(n)$, because 
to build the sequence we only need build the final graph from the initial edge, and saving the resultant graph in each stage, so the 
cost of the algorithm is the number of stage and in the worst of case we will add a node at least in each stage, therefore we have a 
$2n+2$ node, thus the cost of \texttt{BuildX} is $O(n)$.

\subsection{Algorithm \texttt{Aut}}
This algorithm is just do a $O(n)$ times the Algorithm \ref{Aute} and in every stage of \ref{Aute} we will run the algorithms 
\texttt{Ker}, \texttt{Image} and \texttt{Pullback}.

\subsubsection{Algorithm \texttt{Ker}}
The complexity of this algorithm is the complexity of build the function $f: V_{r+1} \rightarrow A_r$ and the complexity of search 
the pairs of nodes who have the same image. The cost of build the function is $O(n)$ assuming that the cost of searching the neighbors
 of a node is $O(1)$, and searching pairs in a vector with $O(n)$ elements is $O(n^2)$, because the vector is not sorted. So 
 the total cost of the algorithm \texttt{Ker} is $O(n^2)$.

 \subsubsection{Algorithm \texttt{Image}}
 The complexity of this algorithm is dominated  of build the sets $A_1, A_2, A'$, the complexity of coloring the set $B_r$ and 
 the complexity of the algorithm \texttt{$C_B(\sigma G)$}. 
 
 The complexity of build the sets $A_1, A_2, A'$ is $O(n^2)$, because for each 
 node in $V(X_r)$ we need to search if it is an ``only child'' or not. The complexity of coloring the set $B_r$ is $O(n^4)$ because the 
 cost of coloring an element of $B_r$ is $O(n)$ and we have $O(n^3)$ elements in $B_r$.
 
 The complexity of \texttt{$C_B(\sigma G)$} is not so easy, we need the complexity of 
 Algorithm \ref{Blocks} and Algorithm \ref{Filter}. The complexity of Algorithm \ref{Blocks} is $O(n^3)$ \cite{LaMi85} and we have seen 
 that Algorithm \ref{Filter} is $O(n^5)O(n^2)$. With all of this we have that the recursive function of the complexity is:
 \[
  T(n) = \left\lbrace \begin{array}{lcl} 1 & if & n = 1 \\ 2 T\left( \frac{n}{2} \right) & if & G \mbox{ is intransitive on } B \\
                       O(n^7) + 4 \left(\frac{n}{2} \right) & if & G \mbox{ is transitive on } B
                      \end{array} \right. 
 \]
so in the worst case 
\[
 T(n) = O(n^7) + 4 \left( \frac{n}{2} \right) = \sum_{i=0}^{\log n}  O \left( 4^i \frac{n^7}{2^i} \right) + 4^{\log n} =O\left( 
 n^8 \right) + n^2 = O( n^8)
\]

Therefore the complexity of \texttt{$C_B( \sigma G)$} is $O \left( n^8 \right)$.
 
 \subsubsection{Algorithm \texttt{Pullback}}

 This algorithm just do the procedure in the Proposition 7, so we only need to extend for $\sigma$ in the generator of the group $S$, which  
 stabilizes the sets $A_1, A_2$ and $A'$ to a $\sigma' \in Aut_e (X_{r+1})$. This extension can be done in $O(n)$ time and we have 
  $O(n)$ generators of $S$, so the cost of the algorithm \texttt{Pullback} is $O(n^2)$.

 Summarizing the complexity of the Algorithm \ref{Aute} is $O(n) \left( O(n^2) + O(n^8) + O(n^2) \right) = O( n^9)$ and the total  complexity
  of the whole  algorithm is $O(n^{10})$
 \section{Improvements for the Implementation}

 We have already seen that we can test if two trivalent graphs are isomorphic in polynomial time, now we will present some improvements of the 
algorithm to show that test if two trivalent graphs are isomorphic  can be do in $O( n^3 \log n)$ time.

\begin{figure}\label{Triplets}\hypertarget{Triplets}{ }
   \centering
    \includegraphics[scale=0.6]{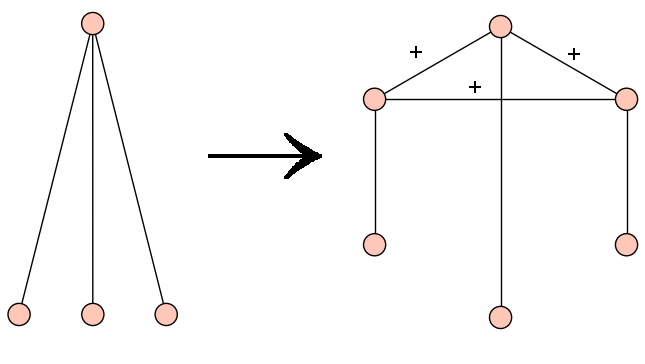}
   \caption{Replacing the triplets in the neighbor sets}
 
\end{figure}

The first improvement is remove the triplets in $A_r$. Recall that triplets are incorporated because the neighbor set of a vertex $v \in V_{r+1}$ 
could have cardinality 3. This situation can be avoid by replacing each such $v$ by a triangle with vertices at ``level'' $r+1$, as we can see 
in \hyperlink{Triplets}{Figure \ref{Triplets}} and having labeled edges. The result is an edge-labeled graph denoted by $\tilde{X}$ with sets 
of size less strict than 3. It is presumed that automorphisms map labeled edges to labeled edges, so the computation of $Aut_e(\tilde{X})$ is
 the same as $Aut_e(X)$ except that $B_r$ need only include the subsets of $V_r$ of size 1 or 2; collection $A_1$ is split into 

\begin{itemize}
 \item[$A_{1a}$] the collection of unlabeled edges connecting vertices in $V_r$. 
 \item[$A_{1b}$] the collection of labeled edges connecting vertices in $V_r$.
\end{itemize}

and an additional color is allowed for an element of $B_r$ to distinguish ether it is in $A_{1a}$, or $A_{1b}$, or neither. 

Also we reformulate $B_r := V_r \times V_r$ in which $(v,v)$ has the color of $v$, while both $(u,v)$ and $(v,u)$ inherit the color of 
$\{u,v\}$. With this color assignment the reassignment retains the identification of $Im(\pi_r)$ with the color preserving subgroup.

It is convenient to present 2-groups in a manner that facilitates several key computations.

\begin{definition}[Smooth generating sequence]\index{Smooth generating sequence}
 Let $G$ be a 2-group generated by $\{g_1, \ldots, g_k \}$, then the sequence $( g_1, \ldots, g_k )$ will be called a \emph{smooth 
 generating sequence} (SGS) for $G$ if $[ G_{(i)} : G_{(i-1)}] \leq 2$, for $i=1, \ldots, ,k$, where $G_{(i)} = \langle g_1, \ldots, 
 g_i \rangle$
\end{definition}

If we have a 2-group $G$ with a smooth generating sequence, then is easy construct an SGS for a subgroup $H$ of index 2.

\begin{lema}
 Let $G$ a 2-group with $ \{ g_1, \ldots, g_k \}$ a SGS, and a subgroup $H$ of index 2. Let $j = \min \{ i \; | \; g_i \notin H \}$ 
 and assign 
 \[
  \tau := g_j 
 \]
 \[
  \beta_i := \left\{ \begin{array}{lcl} g_i & \mbox{ if } & g_i \in H \\ \tau^{-1} g_i & \mbox{ if } & g_i \notin H \end{array} 
  \right. \mbox{ for } i = 1, \ldots, k
 \]

 Then, with $\beta_1, \ldots, \beta_k$ constructed as above
 \begin{enumerate}
  \item $(\beta_1, \ldots, \beta_k )$ is an SGS for $H$.
  \item The time to compute this sequence is $O(k |B|)$, assuming that a membership test requires time $O(|B|)$.
 \end{enumerate}
\end{lema}

\begin{Proof}
 The timing is clear. Let $H_{(i)} = \langle \beta_1 , \ldots, \beta_i \rangle$, then is clear that $H_{(i)} = G_{(i)}$ for all $i < j$ and
 $H_{(i)} \leq G_{(i)}$ for all $i \geq j$. Then, for $i>j$, $g_i \notin G_{(i-1)}$ implies $\beta_i \notin H_{(i-1)}$. 
 So for all $i \neq j$ , $[ H_{(i)} : H_{(i-1)}] \geq [G_{(i)} : G_{(i-1)} ] $. Using that $\{ g_1, \ldots, g_k \}$ is a SGS, we have 
 \[
  \Pi_{i=1}^k [ G_{(i)}: G_{(i-1)}] = | G | = 2 |H| \geq 2 | H_{(k)}| = 1 \Pi_{i=1}^k [ H_{(i)}: H_{(i-1)} ] \geq \Pi_{i=1}^k 
  G_{(i)} : G_{(i-1)} ]
 \]
 
 So, we conclude that $[ H_{(i)}: H_{(i-1)} ] \leq 2$, and $H_{(k)} = H$.
 
\end{Proof}

\begin{remark}
 One can see in \cite{ GaHoLuScWe87} that SGS are preserved through homomorphism and lifting.
\end{remark}

\subsection{Precomputing the Blocks}

The more difficult part of the algorithm is the recursive calls for $C_B( \sigma G)$. The work can be reorganized so as to limit the number of 
distinct blocks, $B$, visited. These blocks form a tree that is precomputed and guides the recursion.

\begin{definition}[Structure tree]\index{Structure tree}
Let $G$ be a 2-group acting on $B$. We call a binary tree $T$ a \emph{structure tree} for $B$ with respect to $G$, $T=Tree(B,G)$, if

\begin{enumerate}
 \item the set of leaves of $T$ is $B$,
 \item the action of any $\sigma \in G$ on $B$ can be lift to an automorphism of $T$.
\end{enumerate}
\end{definition}

It's important to remark, that we can precompute the entire structure tree for the initial $(B,G)$ as follows:

\begin{algorithm}
 
\KwData{$B, G$}
\KwResult{ $T=T(B,G)$}
\Begin{
Let the root of $T$ be $B$ \\
\If{$|B|=1$}{\Return{}}
Find the orbits of $G$ in $B$ \\
\If{$G$ is transitive}{
Find a minimal block system $\{ B_L, B_R \}$ for $G$ on $B$ \\
Find the subgroup $H$ of $G$ that stabilizes $B_L$ \\
Find $\tau \in G \setminus H$ \\
\Return{ $T = Tree(B_L,H) \cup \tau(Tree(B_L, H)$( joined by the new root $B$)}
}
\Else{Partition $B$ into two nontrivial $G$-stable subsets $B_L, B_R$ \\
\Return{$T= T(B_L, G) \cup T(B_R,G)$ ( joined by the new root $B$)}}

}

\end{algorithm}

\begin{lema}
 Given a SGS $(g_1, \ldots, g_k)$ for $G \leq Sym(B)$, $|B|=m$, and let $\Phi(x,y)$ denote the time bound for union-find with 
 $x$ operations on $y$ elements \cite{Aho:1974:DAC:578775}.We have the next time bounds:
 \begin{enumerate}
  \item The orbits of $G$ in $B$ can be computed in time $O(km)$.
  \item If $G^B$ is transitive, a minimal block system $\{ B_L, B_R \}$ for $G$ on $B$ can be computed in time $O( \Phi (2km, 2m))$.
  \item A structure tree $Tree(B,G)$ can be computed in time $O(\Phi(4km,4m))$.
  \item Let $G_r = Aut_e(X_r)$, $B_r = V_r \times V_r$ and $m_r = | V_r|$, then a structure tree $Tree(B_r, G_r)$ can be constructed 
  in time $O( \Phi(4km_r, 4 m_r) + m_r^2 )$. 
  \item The structure trees $Tree(B_r, G_r)$ for all stages, $r=1, \ldots, n-2$ can be constructed in total time $O(n^2)$.
 \end{enumerate}

\end{lema}
A proof of this lemma can be found in \cite{ GaHoLuScWe87}

\subsection{Other improvements}

When we compute $C_B(\sigma G)$, we can avoid deeper recursion, we can change the case 1 where $|B|=1$ by
\begin{itemize}
 \item[Case 1a] $( \exists i \; : \; | B \cap Q_i | \neq | \sigma (B) \cap Q_i | ) : C_B ( \sigma G) := \emptyset$
 \item[Case 1b] $(\exists i \; : \; B \cup \sigma(B) \subseteq Q_i ): C_B( \sigma G) := \sigma G$
\end{itemize}
where $Q_i$ denote the set of elements in $A$ with color $i$.

A non leaf $\tilde{B}$ of $Tree(B,G)$ is called \emph{transitive} if the entry group, $G_{\tilde{B}}$ acts transitively on the set 
$\{ \tilde{B}_L, \tilde{B}_R \}$ and \emph{intransitive}, otherwise. A transitive node $\tilde{B}$ is called \emph{color-transitive}
if the exist group $C_{\tilde{B}} ( G_{\tilde{B}})$, acts transitively on $\{ \tilde{B}_L, \tilde{B}_R \}$. With this definitions 
we can reformulate the conditions in the cases 2 and 3:
\begin{itemize}
 \item[Case 2] ($B$ is intransitive)
 \item[Case 3] ($B$ is transitive)
\end{itemize}

Let $Q= \cup_{i<6} Q_i$, then a node $\tilde{B}$ of $T = Tree(B,G)$ will be called \emph{inactive} if $\tilde{B} \cap Q = \emptyset$ 
and \emph{active} otherwise. We say that the node $\tilde{B}$ is \emph{visited} each time a call to $C_{\tilde{B}}$ does not 
terminate in case 1a and 1b. 

\begin{definition}\index{Pruned tree}
 The subtree $Tree_p(B,G)$ of $Tree(B,G)$, consisting of the active nodes is called the \emph{pruned} tree
\end{definition}

Observe that the pruned tree still guides the recursion.

We call an active node $\tilde{B}$ \emph{facile} if $\tilde{B}$ is intransitive with exactly one active son, and \emph{nonfacile} otherwise. 
Let $\Delta ( \tilde{B} )$ denote the nearest non facile descendant of $\tilde{B}$. Then, if $\sigma$ is color-preserving on $\Delta( \tilde{B})$, 
it must be color-preserving on $\tilde{B}$. Hence, $C_{\tilde{B}} (\tilde{\sigma} G_{\tilde{B}}) = C_{\Delta (\tilde{B})} (\tilde{\sigma} G_{\tilde{B}}) $,
 so that we can pass to node $\Delta(\tilde{B})$. With these facts we have the next algorithm for $C_B( \sigma G)$.
 
 \begin{algorithm}
  \KwData{ $T=Tree(B,G),$ an SGS for $G$}
  \KwResult{ $C_B( \sigma G)$}
  
 \Begin{
\Case{$\exists i : | B \cap Q_i | \neq | \sigma (B) \cap Q_i | ) $}{ \Return{ $\emptyset$}}
\Case{$\exists i : B ºcup \sigma(B) \subseteq Q_i$}{ \Return{ $ \sigma G $}
   }
\Case{ $B$ is facile}{
\Return{$C_{\Delta (B)} ( \sigma G)$}
       }
\Case{$B$ is intransitive}{
\Return{$C_{B_R} C_{B_L} ( \sigma G)$}
}
\Case{$B$ is transitive}{
Find the subgroup $H$ of $G$ that stabilizes $B_L$ \\
Find $\tau \in G \setminus H$\\
\Return{$C_{B_R} C_{B_L} ( \sigma H) \cup C_{B_R} C_{B_L} ( \sigma \tau H)$}
}
}
 \end{algorithm}

\begin{lema}\label{lema9}
 Assuming $Tree(B_r, G_r)$ is constructed as a complete binary tree, adding trivial nodes if it was necessary, it has at most 
 $O(m_r \log m_r)$ active nodes.
\end{lema}

\begin{Proof}
 The pruned tree has at most $2m_r$ leaves. Since $Tree(G_r, B_r)$ has $m_r^2$ leaves, all paths within it, hence all paths within 
 the pruned tree, have length at most $2 \log m_r$.
\end{Proof}

\begin{lema}
 There are at most $2 m_r$ intransitive, nonfacile nodes in the pruned tree
\end{lema}

\begin{Proof}
 Each intransitive, nonfacile node has two sons in the pruned tree, which has $\leq 2 m_r$ leaves.
\end{Proof}

\subsection{The Time Bound}

We know that the structure tree $Tree(B_r, G_r)$ for all $r$ can be found in time $O(n^2)$, and pruning the tree, including the 
construction of $\Delta$, takes $O(n \log n)$. We also need that the entry groups for all nodes of the structure trees and the 
$\tau's$ are computed inf $O(n^3)$ and transitivity is tested for all nodes in time $O(n^2 \log n)$. With all of this we have the 
following theorem.

\begin{theorem}
 Let $X$ be an $n-$vertex, connected, trivalent graph. Then $Aut_e(X)$ can be computed in time $O(n^3)$.
\end{theorem}

A proof of this theorem can be founded in \cite{GaHoLuScWe87}

So we have that the $Aut_e(X)$ can be computed in time $O(n^3)$ and we have a $O(n)$ edges to test, so we derive the following:

\begin{theorem}
 Let $X_1, X_2$ be an $n-$vertex, connected, trivalent graphs. Then test if $X_1$ and $X_2$ are isomorphics can be computed 
 in time $O(n^4)$.
\end{theorem}

This is a great improvement of the first bound that we found in the previous section

\subsection{ More improvements}
In the implementation we made other improvements that don't reduce the theoretical complexity, but they significantly reduce the 
efficient. The improvements are:

\begin{itemize}
 \item Don't compute the whole group $Aut_e(X)$, we only need to know if there is an element of $Aut_e(X)$ that transpose the two 
 elected edges, so we only save the permutations who verify that. It shows especially with large n, when the group $Aut_e(X_r)$ is 
 very large. 
 \item With the previous improvement we stop early in the case that $X_1$ and $X_2$ don't be isomorphic,  because we check every round 
 if there is an isomorphism which exchanges $X_1$ and $X_2$.
 \item Other improvement very useful is check if $\sharp \mathcal{E}(X_1)= \sharp \mathcal{E}(X_2)$. This avoid a lot of computation 
 in the case that $X_1$ and $X_2$ are chosen at random.
 \end{itemize}

\subsection{Other improvements that not be applied}

The theoretical complexity can be improved to $O(n^3 \log n)$, but to this we have calculate previously  the whole group $Aut_e(X_1)$ and
 we can't do the improvements showed previously;  so although the low complexity, the computation time increases. 
 
 Other improvement that not be applied, but it would be useful is the implementation of the own class of permutation group. We 
 note that the most of the time is waste  in the algorithm  while  working with the group of permutations and is the part who more grows 
 when $n$ is bigger. We have two ideas to implement this class:
 \begin{enumerate}
  \item Every permutation is an array of variable size, and when we multiply this permutation with other we only need append two 
  elements to this array. Then to know the image of an element we only need know the position of this element in the array then the 
  image of the element is
 \[
   \sigma (a) = \left\lbrace \begin{array}{lcl} a & if & a \notin \sigma \\ \sigma[i-1] & if & \mbox{the index} i \mbox{ of } a \mbox{ is odd}\\                              \\
                              \sigma[i+1] & if & \mbox{the index} i \mbox{ of } a \mbox{ is pair} 
                              
                             \end{array} \right.
  \]

  remember that in Python the first position in array is the position 0.

This works because in the algorithm we always add a transposition who is disjoint of the previous transpositions.
\item Every permutation is an array of size $n$ and every position show us the image of this position, so the image of $a$ is 
$\sigma[a-1]$.
 \end{enumerate}

\section{General Case}

In this short section, we show  that the trivalent case is extensible to the general case, but we won't depth much as the trivalent 
case,  because the complexity of the algorithm would be too big, although would  keep 
polynomial still. 

We now consider graphs of valence bound by $t$, where $t$ is fixed. It is important to fix $t$ since otherwise the algorithm 
would not be polynomial. The procedure of the trivalent case generalizes, reducing the isomorphism problem to a certain color 
automorphism problem. So, if we show that in the general case $Aut_e(X_r)$ is a 2-group,  then we  provide  the generalization. 

Therefore the reduction to determining the kernel and the image of $\pi_r$ remains intact, the set $A$ now is the collection 
of all non-empty subsets of $\mathcal{V}(X_r)$ of size lower than $t-2$ and the map $f$ has the previous meaning. With all this 
we have that an element $\sigma \in Aut_e (X_{r+1})$ now belongs to the kernel if and only if it stabilizes $f^{-1}(a) $ for 
$a \in A$. These sets form a partition of $\mathcal{V}(X_{r+1}) \setminus \mathcal{V}(X_r)$ and,  $K_r$ is the direct product 
\[
 K_r = \Pi_{a \in A} Sym ( f^{-1}(a))
\]
And each of these factors can be specified with at most two generators, so $K_r$ is a 2-group. We can adapt the proof of 
Proposition \ref{Tutte} and we have that $Aut_e(X_r)$ is a 2-group. Now, using  the rest of the arguments of the trivalent 
case, we get that the general case can be tested in polynomial time. Finally,  $\sigma \in Aut_e(X_r)$ is in the image 
of $\pi_r$ if and only if $\sigma $ stabilizes the sets 
\[
 A_s = \{ a \in A \; | \; f^{-1}(a) = s \} \qquad 0 \leq s \leq t-1
\]
and the set $A'$ of new edges, we need $2t$ colors to color $A$.

\chapter{Implementation test}
 
 Finally,  in this chapter we will present some examples and tests using our own  SAGE implementation. The first examples are to show that the 
 code correctly works, and the test are to prove that  runs in a reasonable time.  Although the SAGE algorithm itself runs more 
 quickly,  they are comparable.
 
 \begin{example}\label{ImpTestEx1}
  In this first example we will test two graphs who are isomorphics. The first graph, is the graph with edges $ \{ (1, 7), (1, 10), 
  (2, 3), (2, 4), (3, 4), (4, 9), (5, 6), (6, 8),$ $ (7, 8), (7, 9),(8, 9) \} $, and the second is the graph with edges $\{ (2, 3), 
  (2, 10), (1, 7), (1, 4),$ $ (7, 4), (4, 9), (5, 6), (6, 8),  (3, 8), (3, 9),(8, 9) \} $, Figure \ref{Example1X1} and \ref{Example1X2} shows this two graphs.

  \begin{figure}
\centering
 \includegraphics[scale=0.5]{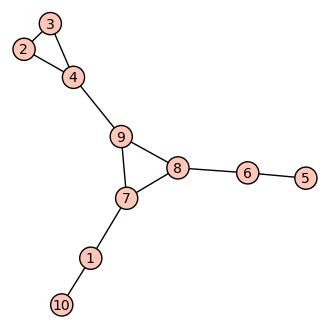}
 \caption{The first graph of Example \ref{ImpTestEx1}}\label{Example1X1}
\end{figure}

\begin{figure}
\centering
\includegraphics[scale=0.5]{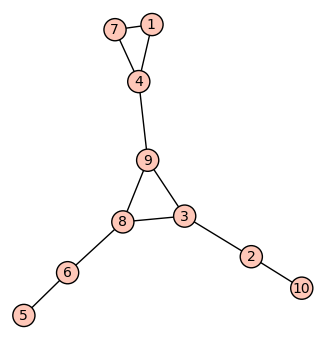}
 \caption{The second graph of Example \ref{ImpTestEx1}}\label{Example1X2}
\end{figure}

The instructions in SAGE for create these graphs area

\begin{verbatim}
 sage: X3=Graph([(1, 7), (1, 10), (2, 3), (2, 4), (3, 4),(4, 9), 
       (5,6),(6, 8), (7, 8), (7, 9),(8, 9)])
 sage: X4=Graph([(2, 3), (2, 10), (1, 7), (1, 4), (7, 4),(4, 9), 
       (5, 6),(6, 8), (3, 8), (3, 9),(8, 9)])
\end{verbatim}

Finally, we test if they are isomorphic: 

\begin{verbatim}
 sage: Isomorphism(X3,X4,10,Iso=True)
1 --> 2
2 --> 1
3 --> 7
4 --> 4
5 --> 5
6 --> 6
7 --> 3
8 --> 8
9 --> 9
10 --> 10
True
\end{verbatim}

Obviously, this produces  an isomorphism between $X_1$ and $X_2$.
\end{example}

\begin{example}\label{ImpTestEx2}

In the following  example we will check two graphs which  are not isomorphic. Figure \ref{Example2X1} and \ref{Example2X2} shows the two graphs 
 to be checked.
\begin{figure}
\centering
 \includegraphics[scale=0.5]{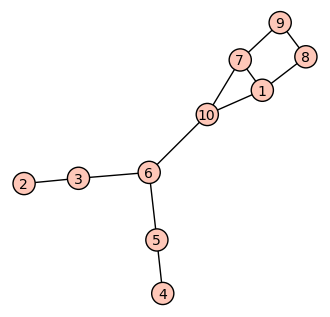}
 \caption{The first graph of Example \ref{ImpTestEx2}}\label{Example2X1}
\end{figure}

\begin{figure}
\centering
\includegraphics[scale=0.5]{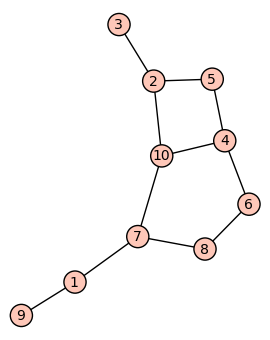}
 \caption{The second graph of Example \ref{ImpTestEx2}}\label{Example2X2}
\end{figure}

The instructions in SAGE for create this graphs area
\begin{verbatim}
sage: X1=Graph([(1, 7), (1, 8), (1, 10), (2, 3), (3, 6),
      (4, 5), (5, 6), (6, 10), (7,9), (7, 10), (8, 9)])
sage: X2=Graph([(1, 7), (1, 9), (2, 3), (2, 5), (2, 10), 
      (4, 5), (4, 6), (4, 10), (6,8), (7, 8), (7, 10)])
sage: Isomorphism(X1,X2,10)
False  
\end{verbatim}

\end{example}

\begin{figure}
\centering
\includegraphics[scale=0.5]{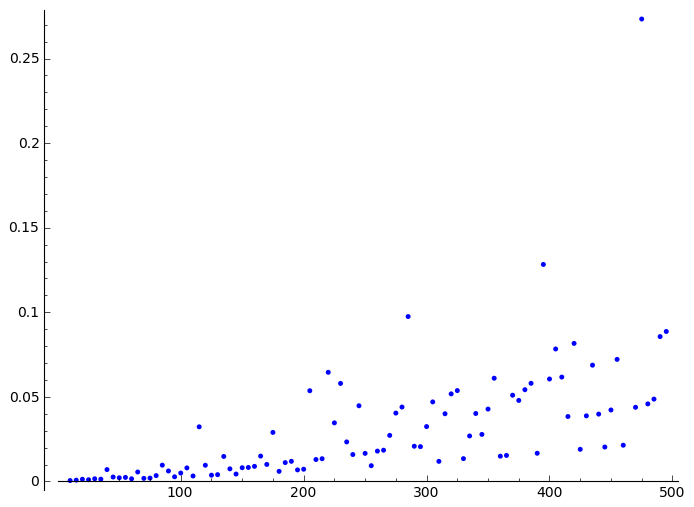}
\caption{Relation seconds-nodes with random graphs}\label{Graphic1}
\end{figure}

Now,  we will present some graphics of different time tests. The first graphic, Figure \ref{Graphic1}, shows the time expend by the 
algorithm to test if two random graphs are isomorphic. The times are so small because if we take two random graphs probably will 
take a different number of edges. Although in the major part of the example the algorithm ends because the graphs have a different 
time of edges, sometimes the algorithm enters in the loop, and in this case the algorithm is relatively efficient.

\begin{figure}
\centering
\includegraphics[scale=0.5]{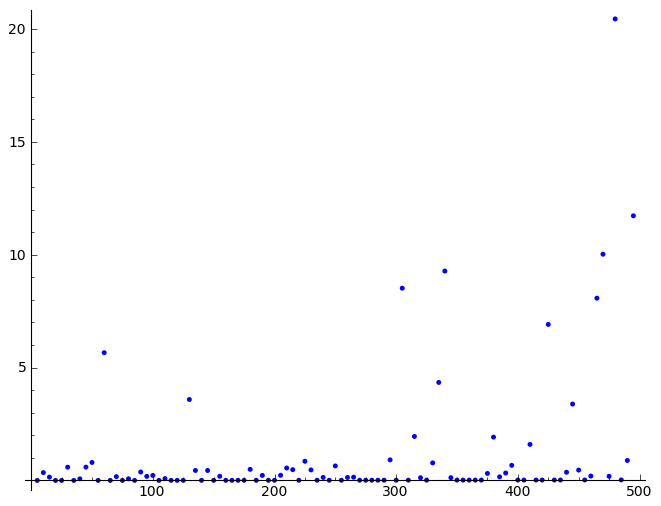}
\caption{Relation seconds-nodes with semirandom graphs}\label{Graphic2}
\end{figure}

\begin{figure}
\centering
\includegraphics[scale=0.5]{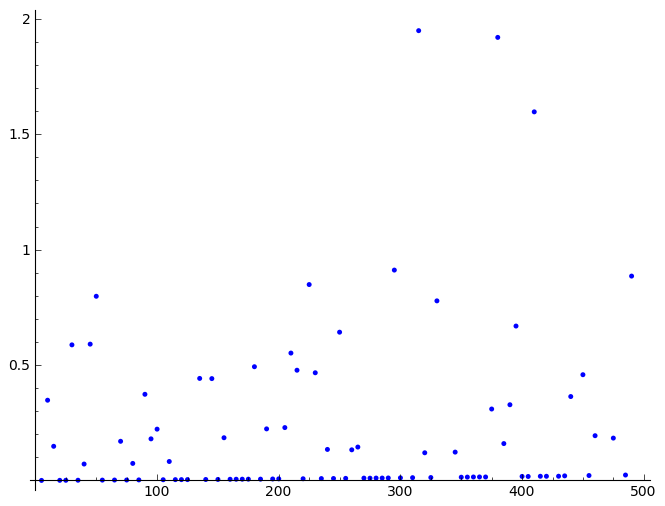}
\caption{Relation seconds-nodes with semirandom graphs, with less than 2 seconds}\label{Graphic3}
\end{figure}

To make the graphs more similar, we perform another test. In this example the degree of the first $n-1$ nodes are the same and the last is 
chosen randomly, this way a third part of the graphs will be isomorphic. In this case,  we also have reasonable times and  the relation 
time-nodes can be seen in Figure \ref{Graphic2} and Figure \ref{Graphic3}

\begin{figure}
\centering
\includegraphics[scale=0.5]{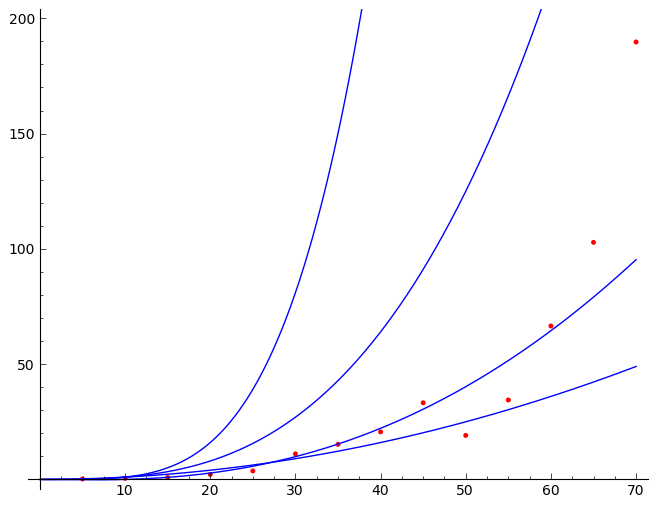}
\caption{Comparison between the algorithm and the functions $(x/10)^4, (x/10)^3, (x/10)^2 \log (x/10), (x/10)^2$ }\label{Graphic4}
\end{figure}

Finally, we will  show what happens if we test isomorphic graphs. In this case the time grows, but we can see in Figure \ref{Graphic4} 
that the time grows more slowly than $(x/10)^3$. The Figure \ref{Graphic4} shows the comparison between the algorithm and the 
functions $(x/10)^4, (x/10)^3, (x/10)^2 \log (x/10), (x/10)^2$

\appendix

\include{n3logn-module}

\chapter*{Sumario}
\addcontentsline{toc}{chapter}{Sumario}

En este trabajo haremos un estudio teórico de un algoritmo para isomorfismo de grafos de valencia acotado propuesto por Eugene 
M. Luks(1982) y una implementación en el sistema SAGE de dicho algoritmo para el caso de valencia 3. 

Este trabajo tiene 4 partes claramente diferenciadas, a saber:
\begin{enumerate}
 \item Preliminares
 \item Algoritmos previos
 \item Algoritmo principal
 \item Pruebas de la implementación
\end{enumerate}

\section*{Preliminares}
En los preliminares tenemos 3 partes: teoría de grupos, teoría de grafos y teoría de la complejidad.

En la primera  presentamos las definiciones básicas de teoría de grupos centrandonos en el grupo de permutaciones, así 
definiciones importantes que se ven son \emph{orbita}, \emph{transitividad}, \emph{G-block} y \emph{G-block system}.

En la segunda, las  definiciones básicas de teoría de grafos, como por ejemplo que és un isomorfismo entre grafos, 
también presentamos algunos resultados, como por ejemplo  que el conjunto de automorfismos de un grafo forman un grupo. 

Finalmente en la tercera y última parte mostraremos conceptos generales  sobre complejidad,  algoritmos polinomiales y una idea intuitiva 
de reducibilidad.

\section*{Algoritmos previos}
En este capítulo presentamos dos tipos de algoritmos, primero veremos algoritmos que se basan en teoría de grupos y luego  
otros dentro de la teoría de grafos.

\subsection*{Algoritmos básicos en teoría de grupos}
Lo más importante y destacable  son los dos lemas siguientes:

\begin{lema}[Furst-Hopcroft-Luks]
Dado un conjunto de generadores para un subgrupo $G$ de $S_n$ se puede determinar en tiempo polinómico
\begin{enumerate}
 \item El orden de $G$.
 \item Saber si una permutación $\sigma$ pertenece a $G$.
 \item Los generadores de un subgrupo de $G$ que sabemos que el índice en $G$ tiene una cota polinomial y,  tenemos un test de 
 pertenencia que se puede ejecutar en tiempo polinomial.
\end{enumerate}
\end{lema}

\begin{lema}
 Dado  un conjunto de generadores para un subgrupo $G$ de $S_n$ y una $G-$orbita $B$, se puede determinar en tiempo polinomial, 
un $G-$block system minimal en $B$. 
\end{lema}

Con el primer lema obtenemos el Algoritmo \ref{alg1} y,  con el segundo obtenemos el Algoritmo \ref{alg2}, que serán importantes en 
el algoritmo principal.

\begin{algorithm}\label{alg1}\hypertarget{alg1}{}
 
\KwData{$\alpha \in G$}
\KwResult{Add $\alpha$ to his $C_i$}
\Begin{
\For{$i \in [0,n-2]$}{
    \If{$\exists \gamma \in C_i \mbox{ : }\gamma^{-1} \alpha \in G_{i+1} $}{
	  $\alpha \leftarrow \gamma^{-1} \alpha $
	}
     \Else{ add $\alpha$ to $C_i$ \\
	    \Return{}
      }
}

\Return{}
}
\caption{Filter}

\end{algorithm}

\begin{algorithm}\label{alg2}\hypertarget{alg2}{}
\KwData{$\omega \neq 1$, $G = \langle g_1, \ldots, g_m \rangle$}
\KwResult{The smallest $G-$block which contains $\{1, \omega \}$}
\Begin{
$C \leftarrow \emptyset$ \\
Set $f( \alpha = \alpha) \; \forall \alpha \in A$ \\
Add $\omega$ to $C$ \\
Set $f(\omega) = 1$ \\
\While{$C$ is nonempty}{
Delete $\beta$ from $C$ \\
$\alpha \leftarrow f( \beta) $\\
$j \leftarrow 0 $ \\
\While{$j < m$}{
$j++$ \\
$\gamma \leftarrow \alpha g_j$ \\
$\delta = \beta g_j$ \\
\If{$f(\gamma) \neq f( \delta)$}{
Ensure $f( \delta) < f(\gamma)$ by interchanging $\gamma$ and $\delta$ if necessary. \\
\For{ $\epsilon \; : \; f( \epsilon ) = f( \gamma )$}{
      Set $f( \epsilon) = f( \delta )$
}
Add $f( \gamma)$ to $C$.
}
}
}
\Return{C}
}
\caption{Smallest $G-$block which contains $\{1, \omega \}$}
\end{algorithm}

\subsection*{Algoritmos básicos en teoría de grafos}
En esta parte se muestra un ejemplo ilustrando que no siempre es un problema complicado el saber si dos grafos son isomorfos. Para 
Mostramos un algoritmo que es $O(n)$ para el isomorfismo de arboles filogenéticos, este  lo presentamos  en Algoritmo \ref{alg3}

\begin{algorithm}\label{alg3}\hypertarget{alg3}{}
 
\KwData{$T_1= (T_1, \rho_1, \{u_1, \ldots, u_2 \}) and T_2=(T_2, \rho_2, \{u_1, \ldots, u_2 \})$}
\KwResult{Test if $T_1$ and $T_2$ are isomorphic}
\SetKwFunction{PostOrderIterator}{PostOrderIterator}
\SetKwFunction{parent}{parent}
\Begin{
Set $\varphi ( \rho_1(u_i)) = \rho_2(u_i) \; \forall i$ \\
Nodes $\leftarrow$ \PostOrderIterator{$T_1$} \\
$w \leftarrow Nodes.next()$ \\
\While{Nodes.hasNext()}{
    \If{ $w$ is not a leaf}{
	\If{ $\varphi (w) == $none}{
	      $v$ child of $w$ \\
	      Set $\varphi(w) = parent( \varphi(v ) )$ 
	    }
	  \For{ $v$ child of $w$}{
	      \If{ $ \varphi(w) \neq $\parent{$\varphi (v) $}}{
		  \Return{False}
		  }
	      }
      }
}

\Return{$\varphi$}
}
\caption{PhylogeneticTreeIsomorphism}
\end{algorithm}

\section*{Algoritmo principal}
En este capítulo veremos el algoritmo principal. La idea general se muestra  en el Algoritmo \ref{alg4}.

\begin{algorithm}\label{alg4}\hypertarget{alg4}{}
\KwData{$X_1, X_2$ connected graphs of bounded valence}
\KwResult{Test if $X_1$ and $X_2$ are isomorphic}
\SetKwFunction{BuildX}{BuildX}
\SetKwFunction{Aut}{Aut}
\Begin{
$e_1 \in \mathcal{E}(X_1)$ \\
\For{$e_2 \in \mathcal{E}(X_2)$}{
	$X \leftarrow $ \BuildX{$X_1, X_2, e_1, e_2 $}\\
	$G \leftarrow $ \Aut($X, e $) \\
	\For{ $\sigma \in G $}{
	    \If{ $\sigma(v_1) == v_2 $}{
		\Return{ True}
	    }
	}
    }

\Return{False}
}
\caption{Isomorphism of graphs of bounded valence}

\end{algorithm}

La estructura de este capítulo esta dividida como sigue:
\begin{itemize}
 \item Valencia 3.
 \item Estudio de la complejidad para el caso de valencia 3.
 \item Mejoras para la implementación.
 \item Generalización al caso general
\end{itemize}
\subsection*{Valencia 3}
En esta parte mostramos como funciona el algoritmo cuando los grafos tienen valencia 3. Para eso,   calculamos  el 
grupo de automorfismos de un grafo, con este fin computamos  una sucesión de grafos y creamos una serie de homomorfismos entre 
los grupos de automorfismos de esa sucesión de grafos. Aqui usaremos el Algoritmo \ref{alg5} y obtendremos la sucesión de 
automorfismos que queríamos.

\begin{algorithm}\label{alg5}\hypertarget{alg5}{}
\KwData{A sequence of graphs $Y$, whose are the result of \texttt{BuildX}}
\KwResult{$Aut_e(X)$ where $X$ is the last graph in the sequence}
\SetKwFunction{Ker}{Ker}
\SetKwFunction{Image}{Image}
\SetKwFunction{Pullback}{Pullback}
\Begin{
$Aut_e = ( e_1 \; e_2 )$
\For{$X \in Y$}{
	$K \leftarrow $ \Ker($X$)\\
	$S \leftarrow $ \Image($Aut_e, X$) \\
	$S2 \leftarrow $ \Pullback($S,X$)\\
	$Aut_e = S2 \cup K$
    }
\Return{$Aut_e$}
}
\caption{The group $Aut_e$}

\end{algorithm}
 
 \subsection*{Estudio de la complejidad}
 En esta parte mostramos de manera más detallada que el algoritmo anterior es polinómico y, que $O(n^{10})$ es una 
 cota superior del coste de dicho algoritmo.
 
 \subsection*{Mejora para la implementación}
Dedicamos esta parte al  estudio de mejoras en vistas de la implementación, estas mejoras serán:
 
 \begin{itemize}
  \item Reducir el tamaño de $A_r$.
  \item Representar los grupos mediante SGS.
  \item Precomputar los bloques.
  \item Otras mejoras.
 \end{itemize}
Con estas mejoras conseguiremos que el algoritmo sea $O(n^4)$, en el peor de los casos.

\subsection*{Caso general}
Finalmente veremos que para el caso general lo único que necesitamos es comprobar que el núcleo de los homomorfismos sigue 
siendo un 2-grupo y, por lo tanto podremos aplicar todo lo demás, adaptándolo para cada valencia.

\section*{Pruebas de la implementación}
Finalmente presentamos  algunos tests realizados con la implementación en el sistema SAGE, con estos 
mostramos  que la cota  superior de $O(n^4)$  no se alcanza y,  que en el caso medio el algoritmo tiene un coste, 
informalmente,  entre  $O(n^3)$ y $O(n^2 \log n)$. 

El apéndice mostramos la documentación de la implementación, aunque se recomienda al lector  
 visitar la pagina {\it http:// www.alumnos.unican.es/aam35/sage-epydoc/index.html} donde hay una  detallada documentación 
en HTML  mucho más fácil y ágil de usar.
\nocite{*}


\bibliography{bibliografia}{}
\bibliographystyle{plain}

\printindex
\end{document}